\documentclass[reprint,aps,prd,superscriptaddress,floatfix,nofootinbib,amsmath,amssymb,showkeys,showpacs]{revtex4-1}
\usepackage{graphicx,color,dsfont,bm}
\usepackage[utf8]{inputenc}
\def\d{\text{d}}
\newcommand{\bra}[1]{\left\langle #1 \right|}
\newcommand{\ket}[1]{\left|#1\right\rangle}
\newcommand{\braket}[2]{\left\langle #1 \right|\left.\! #2 \right\rangle}

\def\cM{\mathcal{M}}
\def\cA{\mathcal{A}}  \def\cAb{\bar{\mathcal{A}}}
\def\bA{\mathbf{A}}  
\def\fA{\mathfrak{A}}  \def\fAb{\bar{\mathfrak{A}}}
\def\bbA{\mathbb{A}}  
\def\cB{\mathcal{B}}  \def\cBb{\bar{\mathcal{B}}}
  
\def\fB{\mathfrak{B}}  
  
\def\cC{\mathcal{C}}  \def\cCb{\bar{\mathcal{C}}}
  
\def\cD{\mathcal{D}}

\def\cN{\mathcal{N}}
\def\cW{\mathcal{W}}
\def\cWI{\overline{\mathcal{W}}}
\def\cWIR{\widetilde{\overline{\mathcal{W}}}}
\def\cWII{\overline{\overline{\mathcal{W}}}}
\def\cWR{\widetilde{\mathcal{W}}}

\def\bt{\mathbf{t}}
\def\bT{\mathbf{T}}
\def\bbf{\mathbf{f}}
\def\pslash{p\!\!\!\slash}
\def\kslash{k\!\!\!\slash}

\def\CF{\mathrm{C_F}}

\def\CA{\mathrm{C_A}}
\def\CAsq{\mathrm{C_A^2}}
\def\CAcub{\mathrm{C_A^3}}
\def\CAfour{\mathrm{C_A^4}}

\newcommand{\mr}[1]{\mathrm{#1}}
\newcommand{\mb}[1]{\mathbf{#1}}
\newcommand{\mc}[1]{\mathcal{#1}}
\newcommand{\qbar}{\bar{q}}
\newcommand{\as}{\alpha_s}
\newcommand{\asb}{\bar{\alpha}_s}
\newcommand{\e}{\epsilon}
\def\X{{\scriptscriptstyle X}}
\def\R{{\scriptscriptstyle\mathrm{R}}}
\def\V{{\scriptscriptstyle\mathrm{V}}}

\newcommand{\tcr}[1]{\textcolor{red}{#1}}

\begin{document}

\title{Eikonal gluon bremsstrahlung at finite $\bm N_c$ beyond two loops}

\author{Yazid \surname{Delenda}}
\affiliation{D\'{e}partement de Physique,  Facult\'{e} des Sciences de la Mati\`{e}re\\
Universit\'{e} Batna 1 - Batna, Algeria}
\email{yazid.delenda@gmail.com}

\author{Kamel \surname{Khelifa-Kerfa}}
\affiliation{D\'{e}partement de Physique, Facult\'{e} des Sciences\\
Universit\'{e} Hassiba Benbouali de Chlef - Chlef, Algeria}
\email{kamel.kkhelifa@gmail.com, k.khelifakerfa@univ-chlef.dz}

\begin{abstract}
We present a general formalism for computing the matrix-element squared for the emission of soft energy-ordered gluons beyond two loops in QCD perturbation theory at finite $N_c$. Our formalism is valid in the eikonal approximation. A {\tt Mathematica} program has been developed for the automated calculation of all real/virtual eikonal squared amplitudes needed at a given loop order. For the purpose of illustration we show the explicit forms of the eikonal squared amplitudes up to the fifth-loop order. In the large-$N_c$ limit our results coincide with those previously reported in literature.
\end{abstract}


\maketitle

\section{Introduction}
\label{sec:Intro}

The huge amount of data collected during the recent runs of the LHC and their analyzes have initiated what one may refer to as ``QCD precision measurements'' programme in analogy to the previous ``Electroweak precision measurements''. For example, the CMS collaboration has recently extracted the QCD coupling $\alpha_s$ with less uncertainties compared to the theoretical predictions \cite{Khachatryan:2014waa}. In fact, recent experimental studies emphasized the indispensability of higher-fixed-order calculations matched with resummed predictions for better description of the data \cite{Chatrchyan:2014gia}. It is therefore essential that more work should be devoted by the phenomenology community to ``QCD precision calculations''. The current paper is a contribution to the latter programme.

It is well known that QCD matrix-element (ME) (or S-matrix) calculations can only be performed within the framework of perturbation theory (PT), which is valid at high energies where QCD becomes asymptotically free. In PT, the matrix element is expanded as a series in the strong coupling parameter $\as$. The calculation of the contributions to the ME at each order is delicate and extremely challenging even at the very first leading orders (see for instance Ref. \cite{Catani:1999ss}).

A particular interesting approximation that has played a crucial role in paving the way to the understanding of the all-orders structure of matrix elements (or scattering amplitudes), and which is implemented in various Monte Carlo parton showers as we shall see below, is the \emph{eikonal} approximation \cite{Levy:1969cr, Abarbanel:1969ek, Wallace:1973iu, Bassetto:1984ik, Mueller:1981ex, Bassetto:1982ma}. In QCD, it is valid in the limit where the momenta of the radiated massless gluons become soft.\footnote{This is the reason why some authors refer to it as the ``soft approximation'', see for example Ref. \cite{Dokshitzer:1991wu}.}  The quark/gluon propagators in particular, and Feynman rules in general, are greatly simplified in the said approximation. The recoil of the radiating energetic particles against the radiated soft gluons may safely be neglected, as we shall see later in the main text when we consider the eikonal Feynman rules. The eikonal approximation is sometimes discussed in terms of \emph{Wilson lines, webs} etc (see for instance Ref. \cite{Laenen:2010uz}).

It has been shown that eikonal amplitudes \emph{exponentiate}, i.e., factorize out into a product of a hard Born amplitude (without radiation) and an exponential containing all radiation, both for abelian \cite{Yennie:1961ad} and non-abelian \cite{Sterman:1981jc, Gatheral:1983cz, Frenkel:1984pz, Catani:1984dp} gauge theories. Schematically, the exponentiation may be written in the case of a non-abelian (QCD) theory for the simple process $e^+ e^- \to q \qbar + g_1  + \cdots + g_n$ as \cite{Laenen:2010uz}:
\begin{equation}
\cM = \cM_0 \, \exp\left(\sum_{i=1}^n \widetilde{\cC}_i\, \widetilde{W}_i \right), \label{eq:EikAmp_exponentiate}
\end{equation}
where $\cM_0$ is the Born amplitude (for the process $e^+ e^- \to q \qbar$), $\widetilde{\cC}_i$ and $\widetilde{W}_i$ are, respectively, the color and kinematical factors of the amplitude in the eikonal limit corresponding to the emission of the $i^{\mr{th}}$ soft gluon. Generalizations of Eq. \eqref{eq:EikAmp_exponentiate} to more complicated QCD processes also exist (see Ref. \cite{Laenen:2010uz} and references therein).

Two important features of the exponential form of the amplitude in Eq. \eqref{eq:EikAmp_exponentiate} are to be noticed: firstly, the computation of the first few orders may reveal great information about the all-orders structure, and secondly, it stands as a backbone to the \emph{resummation} of the large logarithms arising in the calculation of exclusive observable cross-sections in the ``soft regions'' of the phase space. In this regard, the phase space itself must factorize, and hence exponentiate, for the resummation to be at all possible and consequently applicable to QCD processes. Fortunately, this has been demonstrated to be true for various such processes (see Ref. \cite{Laenen:2010uz} for a list of them).

The computation of the exponent in Eq. \eqref{eq:EikAmp_exponentiate} has proven to be quite involved, especially beyond the second order in PT, due to the fact that the color space in QCD is matrix-valued and thus non-commutative in general. Moreover, the number of Feynman diagrams that need to be considered at the $n^{\mr{th}}$ order in PT increases factorially as $(n+1)!$.

It has however been possible for quite a long time to extract the full structure of the emission amplitude squared in the large-$N_c$ limit (with $N_c$ being the degree of the SU($N_c$) group) and for energy-ordered soft gluons \cite{Bassetto:1984ik, Fiorani:1988by}. The large-$N_c$ approximation essentially reduces the matrix-valued color space to simply a scalar-valued space, thus tremendously simplifying the color structure. The strong energy ordering, on the other hand, whereby the energies $\omega_i$ of the soft gluons $k_i$ are ordered as:
\begin{equation}
\omega_{n} \ll \omega_{n-1} \ll \cdots \ll \omega_{1} \ll Q^2\, ,
\label{eq:SEO}
\end{equation}
with $Q$ being the hard scale, is employed to formulate the ``dipole'' picture of gluon bremsstrahlung \cite{Dokshitzer:1991wu}. In this picture, the softest $n^{\mr{th}}$ gluon $k_n$ is emitted by the ensemble of $(n+1)n/2$ harder dipoles. Moreover, the dipole formalism underlines the gluon cascade, implemented in various Monte Carlo algorithms such as the Lund-Dipole-Scheme \cite{Gustafson:1987rq, Gustafson:1986db} and {\tt Ariadne} \cite{Lonnblad:1992tz}, and describes the elementary parton decays (shower) responsible for the structure (and substructure) of final-state jets.

Restoring the full finite-$N_c$ dependence of the exponent in Eq. \eqref{eq:EikAmp_exponentiate}, the progress has so far not been very satisfactory for the reasons mentioned above. The aim of this paper is to show that with the help of the recent developments in the field of computational physics it becomes possible to ``theoretically'' extract the complete analytical form of the exponent (and practically the first few loop orders, due to limited computational power) in Eq. \eqref{eq:EikAmp_exponentiate}. Sj\"{o}dahl developed a {\tt Mathematica} package called {\tt ColorMath} \cite{Sjodahl:2012nk, Sjodahl:2013hra} that automatically performs color-summed calculations in SU$(N_c)$ for arbitrary $N_c \geq 2$. With the aid of this package the issue of the non-abelian color structure that has long jeopardized any progress beyond two loops can straightforwardly be resolved.

It remains thus to determine the kinematic coefficients multiplying the color factors for all possible real/virtual gluon configurations, which represent the set of all possible emitting dipoles at a given order. We have developed a {\tt Mathematica} program to achieve this very goal. We are currently rewriting the program into a form of a package, called {\tt EikAmp}, that should: be able to compute few higher-loop squared amplitudes (ideally up to any loop order), include few callable functions, maximally exploit all available symmetries in order to optimize for minimal execution time and use clear easily-readable syntax. Due to the slowness of {\tt Mathematica} compared to programming languages such as {\tt c++} and {\tt fortran}, we intend to develop alternative versions of {\tt EikAmp} using the said languages in the near future. The aim behind the latter versions of {\tt EikAmp} is not only related to timing but also to make eikonal squared amplitudes readily available for implementation and/or interfacing with other QCD programs, including general-purpose parton showers. It is worthwhile to note that the program can easily be used for QED calculations by simply turning off color matrices. Full details about the program along with its manual will be presented elsewhere.

Note that within the eikonal approximation and assuming strong energy ordering, Catani and Ciafaloni derived in Ref. \cite{Catani:1984dp} an analytic abstract expression for the emission amplitude of $n$ soft (real and/or virtual) gluons to arbitrary orders in PT. Their results were casted into an iterative form which fully accounts for the entire exponentiation of leading soft singularities. They did not, however, perform any explicit calculations, particularly of the color structure of the amplitudes (at least beyond two loops), which had been the chief obstacle for such calculations. Moreover, unlike their amplitudes (squared) which have to be simplified starting from the exponential iterative form, ours are ready for phase-space integration and/or other manipulations without any further simplifications.

Lastly, it is worth mentioning that the results achieved herein have been employed in our recent paper \cite{Khelifa-Kerfa:2015mma} to compute the hemisphere mass distribution in $e^+ e^- \to$ dijet events fully up to four loops and partially at five loops. The distribution was shown to exhibit a pattern of exponentiation both for global (primary abelian emissions) and non-global \cite{Dasgupta:2001sh,Dasgupta:2002bw} (secondary correlated non-abelian emissions) logarithms, and an ansatz for the all-orders resummed result was thus suggested. The eikonal approximation considered in Ref. \cite{Khelifa-Kerfa:2015mma} and throughout this paper only guarantees the resummation of up to single logarithms, of the form $\as^n L^n$ in the exponent of the resummed distribution. To go beyond this accuracy one should consider including the next-to-eikonal corrections presented in Ref. \cite{Laenen:2010uz}. We reserve this work to future publications.

The organization of this paper is as follows. In section \ref{sec:Eikonal_approximation} we present the details of the eikonal approximation at one loop and discuss the corresponding form of the virtual corrections. Employing the eikonal Feynman rules, we derive in section \ref{sec:Eikonal_amplitude_at_nth_loop} the general formula of the eikonal amplitude squared at the $n^{\mr{th}}$ loop order. We discuss in the same section the treatment of virtual gluons at all orders, the implementation of the said formula into {\tt Mathematica} and its reducibility in the large-$N_c$ limit. Section \ref{sec:Examples} is then devoted to presenting the expressions of the eikonal amplitudes squared up to five loops. Finally we conclude in section \ref{sec:Conclusion}.

\section{Eikonal approximation}
\label{sec:Eikonal_approximation}

\subsection{Single real gluon emission}

\begin{figure}[ht]
\centering
\includegraphics[width=0.48\textwidth]{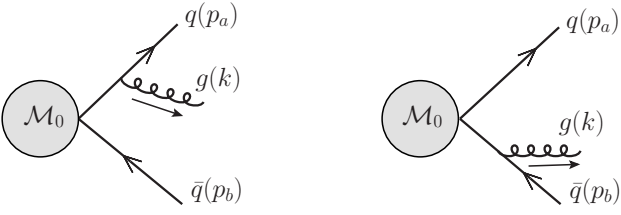}
\caption{\label{fig:EEqqbar+g}The first-order correction to the simple QCD process: electron-positron annihilation into a quark and anti-quark accompanied with the emission of a soft gluon $g$.}
\end{figure}

Consider for simplicity the Feynman diagrams depicted in Fig. \ref{fig:EEqqbar+g}
representing the first-order radiative correction to the hard Born configuration $\cM_0 \to q\qbar$, where $\cM_0$ embodies the leptonic part of the amplitude, and the quark, anti-quark and gluon are on-mass-shell, that is $p_a^2=p_b^2=k^2=0$, with $p_a$, $p_b$ and $k$ the momenta of the quark, anti-quark and gluon respectively. Applying Feynman rules (see for instance Ref. \cite{ellis2003qcd}) one obtains for the amplitude corresponding to the sum of the two diagrams:
\begin{align}
&\imath \cM_1 \left(p_a,p_b,k\right) = \bar{u}_{i}(p_a)\cdot\notag\\
&\cdot\Bigg[\left(-\imath g_s t^c_{ij}\right) \gamma^\nu \e^{c\,\ast}_\nu(k)\frac{\imath\left(\pslash_a+\kslash \right)}{(p_a + k )^2}\, \imath \cM_0 \left(p_a+k , p_b \right) + \notag \\
&+\imath\cM_0 \left(p_a, p_b+k  \right) \frac{\imath \left(\pslash_b+\kslash \right)}{(p_b + k )^2} \left(+\imath g_s t^c_{ij} \right) \gamma^\nu \e^{c\,\ast}_\nu (k ) \Bigg] v_{j}(p_b)\,,
\label{eq:M0+g_Amp}
\end{align}
where $u(p_a)$, $v(p_b)$ and $\e(k)$ are the standard quark and anti-quark Dirac spinors and gluon polarization vector, $g_s$ is the strong coupling, $t^c$ are the color matrices and $\gamma^\nu$ are the Dirac matrices. The eikonal approximation corresponds to the case where the emitted gluon is soft, i.e., when the momentum $|\mb{k}|\ll|\mb{p}_a-\mb{p}_b|$ (or equivalently $k\to 0$ and $p_a\sim p_b$). We can then write $\cM_0\left(p_a+k , p_b\right) \simeq \cM_0\left(p_a, p_b + k \right) \simeq \cM_0\left(p_a, p_b\right)$, neglect the $\kslash $ terms in the numerators of the (anti-)quark propagators in Eq. \eqref{eq:M0+g_Amp} and simplify the squares in the denominators to:
\begin{align}
\left(p_{a(b)} + k  \right)^2 = 2\, p_{a(b)} \cdot k \,.
\end{align}
Using Dirac algebra \cite{ellis2003qcd} one finds for the product of the slashed momenta with the quark and anti-quark spinors:
\begin{subequations}
\begin{align}
\bar{u}(p_a)\gamma^\nu  \pslash_a &= \bar{u}(p_a)\gamma^\nu  \gamma^\mu p_{a\mu} = 2\, p^\nu_a \,\bar{u}(p_a)\,,\\
\pslash_b \gamma^\nu  v(p_b) &= p_{b \mu} \gamma^{\mu} \gamma^\nu  v(p_b) = 2\, p^\nu_b \, v(p_b)\,.
\end{align}
\end{subequations}
Thus Eq. \eqref{eq:M0+g_Amp} reduces to the following \emph{factorized} form:
\begin{align}
\imath \cM_1 \left(p_a, p_b, k \right)  = & \left[\bar{u}_{i}(p_a)\,\imath\cM_0\left(p_a, p_b\right)\,  v_{j}(p_b) \right]\times \notag\\
&\times \left[ g_s \left(\frac{p_a^\nu}{p_a \cdot k}-\frac{p_b^\nu}{p_b \cdot k}\right) t^c_{ij} \,\e^{c\,\ast}_\nu \right].
\label{eq:M0+g_Amp_Factorised}
\end{align}
This is simply the product of the Born amplitude (without bremsstrahlung) with a factor for the emission of a soft gluon. The physical origin of this factorization is the fact that the emitted low-energy (long-wavelength) gluon cannot resolve the underlying harder (shorter-wavelength) subprocess. In fact this factorization property is of paramount importance and can be used iteratively to construct the amplitude of emission of an arbitrary number of gluons, provided they are strongly ordered in energy, as we show in the next section.

The ratio $p^\nu/(p\cdot k)$ in Eq. \eqref{eq:M0+g_Amp_Factorised} defines the effective  Feynman rule for an eikonal emission (combined propagator and vertex). More details are provided in appendix \ref{sec:app:EffectiveFeynmanRules}. It is worth noting that the latter eikonal emission's expression is invariant under rescaling of the emitter's momentum ($p^\mu$). This translates into the fact that one can neglect the recoil effects against soft gluon emissions.

Multiplying the amplitude \eqref{eq:M0+g_Amp_Factorised} by its conjugate, performing the sum over gluon polarizations and simplifying one finds:
\begin{align}
\left|\cM_1 \right|^2 = \mc{B} \times g_s^2\,2\,\CF\, \frac{\left(p_a \cdot p_b \right)}{\left(p_a \cdot k \right) \left(k \cdot p_b \right)}\,,
\label{eq:M0+g_Amp_Factorised-Squared}
\end{align}
with $\mc{B}$ representing the squared Born amplitude summed (averaged) over spins of final-state quarks (initial-state leptons), and $\CF =(N_c^2-1)/(2N_c)$.

\subsection{One-loop virtual correction}
\label{sec:OneLoopVirtualCorr}

\begin{figure}[ht]
\centering
\includegraphics[width=.2\textwidth]{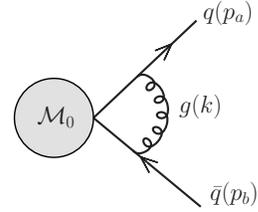}
\caption{\label{fig:EEqqbar+g_OneLoop} One-loop vertex correction to the process $e^+ e^- \to q \qbar$.}
\end{figure}

We shall show in this subsection that, in the eikonal approximation, the squared amplitude for one-loop virtual corrections is simply minus the corresponding real-emission contribution. To this end, consider the Feynman diagram depicted in Fig. \ref{fig:EEqqbar+g_OneLoop},
which represents the vertex correction to the Born amplitude. This is the only contributing diagram at this order since the other two ``self-energy'' diagrams (with a loop starting and ending on the same external leg) result in vanishing contributions in the on-mass-shell limit. The Feynman amplitude for the vertex correction --- in the Feynman gauge --- reads:
\begin{align}
\imath \cM_{\mr{1loop}} = \bar{u}_{i}(p_a)& \bigg[\int \frac{\d^4 k}{(2\pi)^4} \frac{-\imath}{k^2 + \imath \e}\left(-\imath g_s t^{c}_{i\ell}\gamma_{\nu}\right)\cdot\notag\\
&\cdot\frac{\imath\left(\pslash_a + \kslash\right)}{(p_a + k)^2 + \imath \e}\, \imath \cM_0 (p_a + k, p_b - k)\cdot \notag\\
&\cdot\frac{\imath\left(\pslash_b - \kslash\right)}{(p_b - k)^2 + \imath \e}\left(-\imath g_s t^{c}_{\ell j} \gamma^{\nu}\right) \bigg] v_{j}(p_b)\,.
\label{eq:FeynAmp_VerCorr}
\end{align}
Employing the eikonal approximation and simplifying using the Dirac algebra one obtains the factorized expression:
\begin{align}
&\imath \cM_{\mr{1loop}}=\left[\bar{u}_{i}(p_a)\,\imath\cM_0(p_a,p_b)\, \delta_{ij}\,v_{j}(p_b)\right]\times\notag \\
&\times g_s^2 \CF \int \frac{\d^4 k}{(2\pi)^4} \frac{-\imath}{k^2 + \imath \e}\frac{\left(p_a \cdot p_b\right)}{\left(p_a \cdot k + \imath \e \right) \left(- p_b \cdot k + \imath \e \right)}\,.
\label{eq:EikAmp_VerCorr1Loop}
\end{align}
The above integral can be simplified using the techniques of contour integration. Since the integral in the second line of Eq. \eqref{eq:EikAmp_VerCorr1Loop} is manifestly Lorentz invariant we specialize to the center-of-mass (cm) frame of the quark/anti-quark pair, where $\mb{p}_a + \mb{p}_b = 0$, and write:
\begin{align}
\mc{I}
=& \int \frac{\d^4 k}{(2\pi)^{4}}\, \frac{+\imath}{k^2 + \imath \e}\, \frac{\left(p_a \cdot p_b\right)}{\left(p_a \cdot k + \imath \e \right) \left(p_b \cdot k - \imath \e \right) } \notag\\
=& \int \frac{\d^3 \mb{k}}{(2\pi)^3} \int_{-\infty}^{+\infty} \frac{\d k_0}{2\pi}\, \frac{\imath }{\left(k_0 - |\mb{k}| - \imath \e \right) \left(k_0 + |\mb{k}| + \imath \e \right) }\times \notag\\
&\qquad\qquad\qquad\times\frac{2}{\left(k_0 - k_3 +\imath \e \right) \left(k_0 + k_3 - \imath \e \right) }\,,
\label{eq:VerCorr_IntegA}
\end{align}
where we have used for simplicity and without loss of generality:
\begin{subequations}
\begin{align}
p_a&=  \sqrt{s}/2(1,0,0,1)\,,\\
p_b&=  \sqrt{s}/2(1,0,0,-1)\,,\\
k  &=  (k_0,k_1,k_2,k_3) = (k_0,\mb{k}) \,,
\end{align}
\end{subequations}
with $\sqrt{s}$ being the cm energy.\footnote{Recall that we are working in the eikonal limit where recoil of the hard partons is negligible and hence the momenta $p_a$ and $p_b$ are back-to-back.} Note that $p_a \cdot p_b = s/2$. The integrand in Eq. \eqref{eq:VerCorr_IntegA} has four simple poles in the $k_0$ complex plane; two in the upper half and two in the lower half. Closing the contour in the upper half, thus picking up only two poles, and applying the residue theorem one finds:
\begin{equation}
\mc{I}= -\int \frac{\d^3 \mb{k}}{(2\pi)^3}\, \frac{1}{|\mb{k}_\perp|^2} \left(\frac{1}{|\mb{k}|}- \frac{1}{k_3 - \imath \e} \right),
\label{eq:VerCorr_IntegB}
\end{equation}
where $|\mb{k}_\perp|^2 = |\mb{k}|^2 - k_3^2\,$, is the invariant transverse momentum of gluon $k$ in the cm frame of the $q\qbar$ dipole. It is generally given, for the emission of a gluon $k$ off a dipole $(k_i k_j)$, in terms of the \emph{antenna function} $\omega_{ij}$ by the relation:
\begin{equation}
\frac{2}{|\mb{k}_\perp|^2} = \frac{\left(k_i \cdot k_j \right)}{\left(k_i \cdot k \right) \left(k \cdot k_j \right)} \equiv \omega_{ij}(k)\,.
\label{eq:InvTransMom_Dipij}
\end{equation}

The second term in Eq. \eqref{eq:VerCorr_IntegB}, i.e., that involving the $z$-component of the gluon's momentum $k_3$, may be simplified further. Keeping the $\imath \e$ prescription till the end of the evaluation one has \cite{Seymour}:
\begin{align}
& \int \frac{\d^3 \mb{k}}{(2\pi)^3|\mb{k}_\perp|^2}\, \frac{1}{k_3-\imath\e}=\int\frac{\d^2 \mb{k}_\perp}{(2\pi)^3|\mb{k}_\perp|^2}\int_{-\infty}^{+\infty}\d k_3\frac{k_3+\imath \e}{k_3^2+\e^2}\notag \\
&=\int\frac{k_\perp\d k_\perp \d\phi}{(2\pi)^3\,k_\perp^2} \lim_{\xi\to +\infty} \int_{-\xi}^{+\xi} \d k_3 \frac{k_3+\imath \e}{k_3^2+\e^2}\notag\\
&=\frac{1}{(2\pi)^2} \int \frac{\d k_\perp}{k_\perp} \times \imath \pi\,.
\label{eq:VerCorr_IntegC}
\end{align}
Therefore the latter term, which is dubbed ``Coulomb (or Glauber) phase'', is purely imaginary. It is in fact related to \emph{super-leading} logarithms and the possibility of factorization breaking \cite{Forshaw:2006fk, Forshaw:2008cq, Forshaw:2012bi, Angeles-Martinez:2015rna}.

It is worth noting that the said ``$\imath\pi$'' term, which persists at higher loops, cancels when computing physical cross-sections in abelian theories such as QED, while it may have physical consequences in QCD processes involving four or more hard partons. For processes with less than four hard partons such as the process considered herein, that is $e^+e^-\to$ 2 jets,  this phase completely cancels and plays no role to all orders in PT. This can be seen by observing that Coulomb gluons are primary (i.e., directly emitted from the initiating hard dipole $(ab)$) and thus exponentiate, to all orders, into a Sudakov form factor and cancel out in the exponent when multiplied by the conjugate amplitude (see for instance Ref. \cite{Seymour2}).

The final form of the one-loop virtual correction eikonal amplitude reads:
\begin{align}
&\imath \cM_{\mr{1loop}} =\left[\bar{u}_{i}(p_a)\,\imath\cM_0(p_a,p_b)\, \delta_{ij}\,v_{j}(p_b)\right] \times g_s^2 \,\CF \times\notag\\
&\times\left[-\int \frac{\d^3 \mb{k}}{(2\pi)^3\, 2 |\mb{k}|} \frac{ \left(p_a \cdot p_b\right)}{\left(p_a \cdot k \right) \left(k \cdot p_b\right)}
+\frac{1}{(2\pi)^2}\,\imath \pi \int \frac{\d k_\perp}{ k_\perp} \right]\label{eq:EikAmp_1Loopsemifinal}.
\end{align}
To compute the virtual-emission \emph{squared} amplitude we first multiply the above result by the conjugate Born amplitude $[\bar{u}_{\ell}(p_a)\,\imath\cM_{0}(p_a,p_b)\, \delta_{\ell m}\,v_{m}(p_b)]^*$, and then add the complex conjugate of the result, i.e., we add the product of the Born amplitude and the conjugate  of the one-loop amplitude \eqref{eq:EikAmp_1Loopsemifinal}. Doing so we note that the Coulomb phase \eqref{eq:VerCorr_IntegC} disappears in the sum, and we finally obtain:
\begin{align}\label{eq:EikAmp_1Loopsemifinal2}
|\cM_{\mr{1loop}}|^2 = \mc{B} \left[-g_s^2\,2\,\CF\int \frac{\d^3 \mb{k}}{(2\pi)^3 2 |\mb{k}|}\, \frac{\left(p_a \cdot p_b\right)}{\left(p_a \cdot k \right) \left(k \cdot p_b\right)}\right],
\end{align}
which is minus the real-emission amplitude squared \eqref{eq:M0+g_Amp_Factorised-Squared} (after the inclusion of the appropriate phase-space factor), as stated earlier. This result has also been arrived at in Ref. \cite{Schwartz:2014wha}.

This useful consequence of the eikonal approximation will be employed in the next section to deduce eikonal squared amplitudes for configurations in which the softest gluon is virtual, at any order, without explicit calculations. Since real and virtual (softest-gluon) contributions are equal up to a sign, we shall refer to a given order $\as^n$ in the PT expansion as the $n^{\mr{th}}$ loop order (or simply $n$-loop order), although formally the terminology ``loop'' is reserved for virtual corrections.

\section{Eikonal squared amplitudes}
\label{sec:Eikonal_amplitude_at_nth_loop}

\subsection{General form}

Having discussed the eikonal approximation by explicitly carrying out analytical leading-order calculations and demonstrated that: a) the emission amplitude squared factorizes out into a Born amplitude squared multiplying a color factor and a dipole antenna function, and b) virtual corrections are identical to real contributions up to a sign, we are in a position to utilize these findings to build up a generalized form of eikonal amplitudes squared, which is valid at any order in PT expansion.

We begin by considering the (tree-level) emission amplitude of a soft gluon $k_m$ by an ensemble of: quark $p_a$, anti-quark $p_b$ and $(m-1)$ energy-ordered (see Eq. \eqref{eq:SEO}) harder gluons $k_{i=1,\hdots ,m-1}$ \cite{Dokshitzer:1991wu, Forshaw:2008cq}:
\begin{align}
\ket{m} = g_s \sum_{i\in U_{m-1}} \frac{k_i \cdot \e^{a_{m}\,\ast}}{k_i \cdot k_{m}}\, \bT^{a_{m}}_i \ket{m-1},
\label{eq:EikAmp_general}
\end{align}
where $U_{m}=\{a,b,1,2,\hdots,m\}$ is the set of all possible emitters of the $(m+1)^{\mr{th}}$ gluon such that $U_0=\{a,b\}$, $k_a=p_a$ and $k_b=p_b$. The superscript $a_{m}$ is the color index of the emission $k_{m}$ and the ket $\ket{m-1}$ is an $(m+1)$-dimensional vector in color space representing the $(m+1)$-parton amplitude prior to the emission of the soft gluon $k_m$.\footnote{There are $(m-1)$ gluons in addition to the quark and anti-quark pair, thus resulting in $(m+1)$-dimensional vector.} The set of basis kets spanning the color space at a given order $m$, $\ket{c_a, c_b, c_1, \hdots, c_m}$, where $c_{a,b}= 1,\hdots,N_c$ denote the color indices of the quark and anti-quark and $c_{i=1,\hdots,m}= 1,\hdots,N_c^2-1$ denote the color indices of the gluons, may be taken to form an orthonormal set representing the various color states of all $(m+2)$ outgoing partons. In this notation, the amplitude for the production of $(m+2)$ partons in the final state with colors $c_a,c_b,c_1,\hdots,c_m$, is given by \cite{Catani:1996vz,Catani:1996jh}:
\begin{align}
\mathcal{M}_{m}^{c_a,c_b,c_1,\hdots,c_m} = \braket{c_a,c_b,c_1,\hdots,c_m}{m}\,,
\end{align}
and the squared amplitude, summed over final-state colors, is given by:
\begin{align}
\left|\cM_m \left(p_a, p_b, k_1, \hdots,k_m \right) \right|^2 = \braket{m}{m}\,.
\end{align}

Eq. \eqref{eq:EikAmp_general} easily follows from the effective Feynman rules (see appendix  \ref{sec:app:EffectiveFeynmanRules}) as well as the factorization property of eikonal amplitudes discussed in the previous section. The color operator $\bT^{a_m}_i$ in Eq. \eqref{eq:EikAmp_general} acts in the color subspace of parton $i$ and plays the role of inserting a gluon whose color index is $a_m$ and ``repainting'' the color of parton $i$. Thus when acting on the $(m+1)$-dimensional basis vector the operator $\bT^{c}_i$ produces an $(m+2)$-dimensional vector:
\begin{multline}
\bT^c_i \ket{c_a, c_b, c_1, \hdots, c_i, \hdots, c_{m-1}} = \\
= T^c_{c'_i c_i} \ket{c_a, c_b, c_1, \hdots, c'_i, \hdots, c_{m-1}, c }, \label{eq:ColOperator_Action}
\end{multline}
where:
\begin{itemize}
\item $T^c_{c'_i c_i} = t^c_{c'_i c_i}\equiv \bt^c_i$ if the emitting parton $i$ is a quark.
\item $T^c_{c'_i c_i} = - t^c_{c_ic'_i} \equiv - \bt^c_i$ if the emitter is an anti-quark.
\item $T^c_{c'_i c_i} = i f^{c'_i c c_i} \equiv \imath  \bbf^c_i$ if the emitter is a gluon.
\end{itemize}
The sign conventions for the eikonal vertices are presented in appendix \ref{sec:app:EffectiveFeynmanRules}. The operator $(\bT^{a_m}_i)^\dagger$ plays the opposite role to that of $\bT^{a_m}_i$, i.e., when acting on an $(m+2)$-dimensional basis vector it produces an $(m+1)$-dimensional basis vector as it removes one gluon. Moreover, two important identities are to be noted for the operators $\bT^{a_m}_i$ \cite{Catani:1996vz,Catani:1996jh}:
\begin{equation} \label{eq:ColorOperators_Identities}
\begin{split}
&\sum_{i\in U_{m-1}} \bT^{a_m}_i \ket{m-1} = 0\,,\quad\mr{and}\\
& \bT^{a_m}_i \cdot \bT^{a_m}_j = \bT^{a_m}_j \cdot \bT^{a_m}_i; \qquad  \left(\bT_i^{a_m}\right)^2 = C_i \,,
\end{split}
\end{equation}
the first of which demonstrates that the $(m+1)$-parton amplitude transforms as a color singlet under SU(3), and the second shows that the product of two such operators, which shows up when computing squared amplitudes, is invariant under the exchange of the emitter and ``absorber'' of the softest gluon. In other words, the color factors associated with the dipoles $(ij)$ and $(ji)$ are identical. The factor $C_i$ is the Casimir scalar, which is equal to $\CA = N_c$ if $i$ is a gluon, and $\CF$ if $i$ is a quark or anti-quark. Worth mentioning is that the color operators in the fundamental representation $\bt^a$ are both traceless and hermitian, while in the adjoint representation $\imath \bbf^{a}$ are constants (known as ``structure constants'').

Iterating the amplitude \eqref{eq:EikAmp_general} down to the Born level, $\ket{0}$, one has:
\begin{align}
\ket{m}  = & g_s^{m}\sum_{i\in U_{m-1}} \frac{k_{i} \cdot \e^{a_{m}\ast}}{k_i \cdot k_{m}}\bT^{a_{m}}_i  \sum_{j\in U_{m-2}} \frac{k_j \cdot \e^{a_{m-1}\ast}}{k_j \cdot k_{m-1}} \bT^{a_{m-1}}_j  \cdot \notag\\
&\cdots \sum_{\ell\in U_{0}} \frac{k_\ell \cdot \e^{a_{1}\ast}}{k_\ell \cdot k_{1}} \bT^{a_1}_\ell \ket{0} \notag \\
=& g_s^{m} \left(\prod_{n=1}^m  \sum_{i_n\in U_{n-1}} \frac{k_{i_n} \cdot \e^{a_{n}\ast}}{k_{i_{n}} \cdot k_{n}}\right) \times \notag\\
&\times \bT^{a_{m}}_{i_m} \bT^{a_{m-1}}_{i_{m-1}} \cdots \bT^{a_1}_{i_1} \ket{0},
\label{eq:EikAmp_Ket-m+1-to_Ket-2}
\end{align}
where the kinematic factors are scalars and have thus been pulled to the left leaving only the ordered non-commutative color operators to the right. It is worth noting that the ordering of the color operators $\bT^{a_n}_i$ may not necessarily be the same as the ordering of the resultant color matrices $\bt^{a_n}_i$.

\begin{figure}[ht]
\centering
\includegraphics[scale=.9]{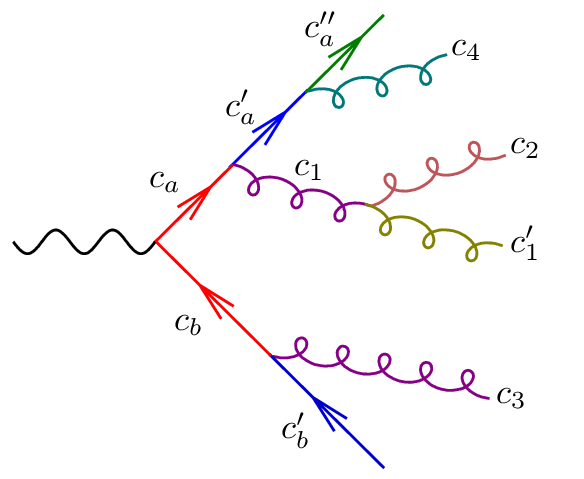}
\caption{Feynman diagram for one of the contributions to the emission amplitude of four gluons in the process $e^+ e^- \to q \bar{q}$.\label{fig:ColorOperation}}
\end{figure}

To see this, consider for example the computation of one of the contributions to the emission amplitude of four gluons (focusing only on the color part and ignoring the kinematic part) as illustrated in Fig. \ref{fig:ColorOperation}.
The corresponding amplitude reads:
\begin{align}
\widetilde{\ket{4}} &= \delta_{c_ac_b}\,\bT^{c_4}_a \, \bT^{c_3}_b \, \bT^{c_2}_1 \, \bT^{c_1}_a \ket{c_a, c_b}\notag\\
&= t^{c_1}_{c'_a c_a} \delta_{c_ac_b}  \bT^{c_4}_a \, \bT^{c_3}_b\, \bT^{c_2}_1  \ket{c'_a, c_b, c_1}\notag\\
&= t^{c_1}_{c'_a c_a} \delta_{c_ac_b} \imath f^{c_2}_{c_1' c_1}  \bT^{c_4}_a\, \bT^{c_3}_b \ket{c'_a, c_b, c_1', c_2} \notag\\
&= t^{c_1}_{c'_a c_a} \delta_{c_ac_b} (-t^{c_3}_{c_b c_b'})\, \imath f^{c_2}_{c_1' c_1} \bT^{c_4}_a \ket{c'_a, c_b', c_1', c_2, c_3}\notag\\
&= -t^{c_4}_{c_a'' c_a'} t^{c_1}_{c'_a c_a} \delta_{c_ac_b} t^{c_3}_{c_b c_b'} \imath f^{c_2}_{c_1' c_1}   \ket{c''_a, c'_b, c_1', c_2, c_3, c_4}\notag\\
&= - \left(\bt^{c_4}_a\, \bt^{c_1}_a\, \bt^{c_3}_b\right)_{c_a''c_b'}  \left(\imath \bbf^{c_2}_1\right)_{c_1c_1'} \ket{c''_a, c'_b, c_1', c_2, c_3, c_4},
\end{align}
where we made use of the fact that $\ket{0}\propto\delta_{c_ac_b}\ket{c_a,c_b}$. Thus the ordering of the matrices $\bt^a$ is different from that of the operators $\bT^a$.

The corresponding conjugate (bra) amplitude $\bra{m}$ for real emissions reads:
\begin{align}
\bra{m} =& g_s^{m} \left(\prod_{n=1}^m  \sum_{j_n\in U_{n-1}} \frac{k_{j_n} \cdot \e^{a_{n}}}{k_{j_n} \cdot k_{n}}\right) \times\notag\\
&\times\bra{0} \left(\bT^{a_1}_{j_1}\right)^\dagger \cdots \left(\bT^{a_{m-1}}_{j_{m-1}} \right)^\dagger  \left(\bT^{a_{m}}_{j_m}\right)^\dagger.
\label{eq:EikAmp_Bra-m+1-to_Bra-2}
\end{align}
The eikonal amplitude squared, summed over gluon polarizations, is then given by:
\begin{align}
\braket{m}{m} =& \left(-1\right)^{m} g_s^{2m}\left(\prod_{n=1}^m\, \sum_{i_n\neq j_n\in U_{n-1}} \omega_{i_n j_n}(k_n) \right)\times \notag\\
&\times\bra{0} \left(\bT^{a_1}_{j_1}\right)^\dag \cdots \left(\bT^{a_{m}}_{j_{m}}\right)^\dag \bT^{a_{m}}_{i_{m}}  \cdots  \bT^{a_1}_{i_1} \ket{0},\label{eq:EikAmpSq_A}
\end{align}
where the factor $(-1)^{m}$ results from the sum over polarizations for each gluon, $\omega_{ij}(k)$ is the antenna function defined previously, Eq. \eqref{eq:InvTransMom_Dipij}, and we have excluded all terms with indices $i_n=j_n$ in the sum since they vanish by the on-mass-shell condition $k_{i}^2 = 0$ for $i=a,b,1,\hdots,m$. Eq. \eqref{eq:EikAmpSq_A} may be recast, for the emission of $m$ real gluons, in the form:
\begin{equation}
\braket{m}{m}= \braket{0}{0}  W_m \,,
\label{eq:EikAmpSq_B}
\end{equation}
where $W_m$ is the \emph{factorized} eikonal amplitude squared for $m$ soft energy-ordered gluons in the final state:
\begin{align}
&W_m = \left(-1\right)^m g_s^{2m} \left(\prod_{n=1}^m \, \sum_{i_n\neq j_n\in U_{n-1}} \omega_{i_n j_n}(k_n) \right)\times\notag\\
&\times \frac{1}{N_c} \bra{c_a',c_a'} \left(\bT^{a_1}_{j_1}\right)^\dag \cdots \left(\bT^{a_{m}}_{j_{m}}\right)^\dag \bT^{a_{m}}_{i_{m}}  \cdots  \bT^{a_1}_{i_1} \ket{c_a,c_a},
\label{eq:EikAmpSq_mGluon}
\end{align}
where the Born amplitude for the simple $e^+e^- \to q\qbar$ process is proportional to $\delta_{c_ac_b}\ket{c_a,c_b} = \ket{c_a,c_a}$. Thus a factor $N_c = \braket{c_a',c_a'}{c_a,c_a}= \left(\delta_{c_a c_a'}\right)^2$ has been divided by in Eq. \eqref{eq:EikAmpSq_mGluon}. Doing so one achieves a complete factorization of $W_m$ from the Born contribution $\mathcal{B}=\braket{0}{0}$.

The differential cross-section for the emission of $m$ energy-ordered soft (real) gluons, normalized to the Born cross-section $\sigma_0$, may then be written in the following form:
\begin{align}
\frac{\d \sigma_m}{\sigma_0} = \d \Phi_m \, W_m \,,  \label{eq:EikCrossX}
\end{align}
where the phase-space factor reads:
\begin{align}
\d\Phi_m =& \prod_{i=1}^m \frac{\d^3 \mb{k}_i}{(2\pi)^3 \,2\, \omega_i} = \prod_{i=1}^m \frac{\omega_i \, \d \omega_i}{4\pi^2} \, \frac{\d\Omega_i}{4\pi}\notag\\
=& \prod_{i=1}^m \frac{k_{ti} \d k_{ti}}{4\pi^2}\, \frac{\d\eta_i\, \d\phi_i}{4\pi} \,,
\label{eq:EikPhaseSpaceFactor}
\end{align}
with $\omega_i$ and $k_{ti}$ the energy and transverse momentum of the $i^{\mr{th}}$ gluon, $\d\Omega_i$ its differential solid angle, and  $\eta_i$ and $\phi_i$ are, respectively, its rapidity and azimuthal angle. We have $k_{ti}=\omega_i\sin\theta_i$ and $\eta_i=-\ln\tan(\theta_i/2)\,$, with $\theta_i$ the polar angle of gluon $k_i$. It is worth mentioning that the above differential cross-section for $m$ gluons in the final state must be symmetric under the exchange of the indistinguishable gluons (bosons). This means that one has to sum over all possible permutations of gluons and divide the result by $m!$. This is, however, equivalent to considering only a specific ordering of the gluons and omitting the $1/m!$.

Substituting the amplitude squared \eqref{eq:EikAmpSq_mGluon} into the expression of the eikonal differential cross-section \eqref{eq:EikCrossX}, and including all virtual corrections, the \emph{total} cross-section may be cast in the following final form:
\begin{align}
\frac{\d\sigma}{\sigma_0} = \sum_{m=1} \asb^m \left(\prod_{i=1}^m \frac{\d k_{ti}}{k_{ti}}\,\d\eta_i\, \frac{\d\phi_i}{4\pi}\right) \sum_{\X} \cW^\X_{12\hdots m} \,,
\label{eq:EikCrossX_Final}
\end{align}
where $\asb = \as/\pi=g_s^2/(4\pi^2)\,$, and the second sum (over $X$) is over all possible real/virtual gluon configurations. In other words, $X =\mr{x_1 x_2 \hdots x_m}\,$, where $\mr{x_i} \in \{\mr{R}, \mr{V}\}$ corresponds to whether the $i^{\mr{th}}$ gluon is real (R) or virtual (V). Virtual corrections in the expression \eqref{eq:EikCrossX_Final} have been arrived at along the same lines as the real-emission contributions, including the factorization of the phase space which also follows from the Bloch-Nordsieck theorem \cite{Bloch:1937pw}. Since real and virtual soft contributions exactly cancel out in the total cross-section (of sufficiently-inclusive observables) one can therefore express the virtual contributions as integrals over momenta of exactly the same form as real emissions, which leads to a factorization of the virtual phase space that is identical to that of real emissions as shown in Eq. \eqref{eq:EikCrossX_Final}. Further details about the virtual corrections are presented in the next subsection \ref{sec:Treatment of virtual gluons}.

The final form of the eikonal amplitude squared $\cW_{12\hdots m}^\X$ for the emission of $m$ energy-ordered soft gluons in a given configuration $X$ reads:
\begin{align}
\cW_{12\hdots m}^\X &= \left(\prod_{n=1}^m \, \sum_{i_{n} \neq j_{n}\in U_{n-1}} w^{n}_{i_{n} j_{n}}\right) \cC^{i_1 i_2 \hdots i_{m}}_{j_1 j_2\hdots j_{m}} \,, \label{eq:EikAmpSq_Final}
\end{align}
where we have defined the purely-angular antenna function $w^\ell_{ij} = k_{t\ell}^2\, \omega_{ij}(k_\ell)$ (or alternatively $w^\ell_{ij} = \omega_{\ell}^2\, \omega_{ij}(k_\ell)$ if one uses energies and polar angles instead of transverse momenta and rapidities in the phase space \eqref{eq:EikPhaseSpaceFactor}) and introduced the color factor:
\begin{multline}
\cC^{i_1 \hdots i_{m}}_{j_1\hdots j_{m}} = \frac{(-1)^m}{N_c}\times\\
\times \bra{c_a',c_a'}\left(\bT^{a_1}_{j_1}\right)^\dag \cdots\left(\bT^{a_m}_{j_m}\right)^\dag \bT^{a_{m}}_{i_{m}}\cdots\bT^{a_1}_{i_1} \ket{c_a,c_a}. \label{eq:EikColorFactor}
\end{multline}
Though the general form of the above color factor holds true for the configuration in which all gluons are real, there are few changes that take place when one or more gluons are virtual. We discuss this issue in the next subsection. The factor $(-1)^m$ ensures that the color factor $\cC$, and hence the amplitude squared \eqref{eq:EikAmpSq_Final}, is positive for all values of $m$ (for all-gluons-real configurations).

Few important properties of the eikonal amplitude squared \eqref{eq:EikAmpSq_Final} are in order:
\begin{itemize}
\item It is totally symmetric under the interchange of the two hardest partons (the quark and anti-quark $a \leftrightarrow b$). For the hardest gluon, we shall thus only consider one emitting dipole, say $w_{ab}^{1}\,$, ignore $w_{ba}^{1}$ (i.e. consider only $i_1=a$ and $j_1=b$) and multiply the final result by a factor of $2$. This will prove useful in decreasing the execution time of the {\tt EikAmp} program.

\item It is also symmetric under the interchange of the legs of the dipole emitting the next-to-hardest gluon ($i_2\leftrightarrow j_2$) as well as the softest gluon ($i_m\leftrightarrow j_m$). Hence we pick up one ordering for the said dipole legs (say $i_2>j_2$ and $i_m>j_m$) and multiply the final result by a factor of $4$.

\item For the softest gluon, it is always true that:
    \begin{align}
    \cW^{\mr{x}_1 \hdots \V}_{12 \hdots m} = -  \cW^{\mr{x}_1 \hdots \R}_{12 \hdots m}\,,
    \label{eq:EikAmp_RV_Sym}
    \end{align}
    regardless of the nature, real or virtual, of the rest of the gluons. This equation is a generalization of the one-loop result \eqref{eq:EikAmp_1Loopsemifinal2} to arbitrary orders in PT, and can straightforwardly be deduced by iterating the procedure outlined in subsection \ref{sec:OneLoopVirtualCorr}. Since Eq. \eqref{eq:EikAmp_1Loopsemifinal2} has been arrived at with no explicit role being played by the Born term $\cM_0$ in the integration over $k$, one can simply, for each Feynman diagram involving a softest virtual gluon $k_m$, absorb all harder gluons into a ``Born-like'' amplitude $\cM_{m-1}$ and perform the integration over the momentum $k_m$ in exactly the same way as in eq. \eqref{eq:VerCorr_IntegA}. The relative minus sign between the real and virtual contributions is due to the absence of the sum over the virtual-gluon polarization. The expression \eqref{eq:EikAmp_RV_Sym} also suggests that each virtual diagram corresponds to exactly the same color factor as the corresponding real-emission diagram (just like at one loop where both had the color factor $\CF$). This result can straightforwardly be verified using the {\texttt{EikAmp}} program for any given diagram and at any loop order.

    This observation will serve to reduce the number of squared amplitudes that need to be calculated at a given loop order to a half, thus saving substantial amount of work and time.

\item Since the hardest gluon $k_1$ may only be radiated by the hardest dipole $(ab)$ --- due to strong energy ordering --- the color factor \eqref{eq:EikColorFactor} will always be proportional to the Casimir scalar (color charge) $\CF$. In other words, the color factor at a given loop order $L$ will be proportional to $\left(\CF\right)^n\left(\CA\right)^{L - n}$, with $n=1,\hdots, L$.

\item \emph{Proper} finite-$N_c$ contributions to the squared amplitude \eqref{eq:EikAmpSq_Final} first appear at four loops and have a color factor that is proportional to $(\CF-\CA/2)$. Higher-loop finite-$N_c$ contributions are \emph{always} proportional to the latter color factor.

\item At and beyond four loops, the eikonal amplitude squared for a given configuration $X$ may be split into two contributions; large and finite $N_c$:
    \begin{align}
    \cW^\X_{12\hdots m} = \cW^{\X,\, \mr{lN_c}}_{12\hdots m} + \cW^{\X,\,\mr{fN_c}}_{12\hdots m}\,.
    \label{eq:EikAmp_L-F_nc_Split}
    \end{align}
    Whilst the large-$N_c$ contribution has been known for a while (see for instance Refs. \cite{Bassetto:1984ik, Fiorani:1988by, Banfi:2002hw}), the finite-$N_c$ contribution has not previously been properly addressed in the literature. The present work serves to shed some light on this very contribution.

\item The said finite-$N_c$ contributions to $\cW^{\X}_{12\hdots m}$ have some peculiar features that are absent in the large-$N_c$ contributions. For instance, they are not symmetric under the interchange of the emitted gluons $k_1, k_2, \hdots, k_m$. Moreover, they are not symmetric under the interchange of the legs of each and every single emitting dipole. These symmetry breakings are primarily due to the associated color factor (of non-planar diagrams). It is therefore incorrect, at least at and beyond four loops, to set $i_n > j_n$ in Eq. \eqref{eq:EikAmpSq_Final} and multiply by $2$ for each dipole, resulting in a total common factor of $2^m$. It remains, however, true to set $i_n>j_n$ for $n=1,2,m$, i.e. for the legs of the dipoles emitting the hardest, next-to-hardest and softest gluons, and multiply by a factor 8, as discussed above.
\end{itemize}

More details on the properties of the eikonal amplitude squared \eqref{eq:EikAmpSq_Final} will be presented in the next section \ref{sec:Examples}. Before that, we first discuss the form of the color factor \eqref{eq:EikColorFactor} in the case where one or more gluons are virtual, succinctly describe the {\it Mathematica} program utilized to compute the squared amplitudes presented in Sec. \ref{sec:Examples} and briefly illustrate how the amplitude squared \eqref{eq:EikAmpSq_Final} reduces in the large-$N_c$ limit.

\subsection{Treatment of virtual gluons}
\label{sec:Treatment of virtual gluons}

The calculation of the amplitude squared for the case of emission of virtual gluons (other than the softest) requires a special attention. Firstly we note that the aforementioned equality \eqref{eq:EikAmp_RV_Sym} is --- as clearly stated --- only valid for deducing the squared amplitude when the \emph{softest} gluon is made virtual, without altering the real/virtual configuration of the other gluons, and is generally not applicable for the case when arbitrarily harder gluons are virtual. This is mainly due to the fact that the ordering of color operators changes as gluons are made virtual for a given Feynman diagram.

Consider, for instance, a virtual gluon $k_\ell$ attaching the legs of a (real) dipole $(i_\ell j_\ell)$. Writing the product of ordered color operators proceeds in an ordinary way (i.e., starting from the hardest-gluon color operator $\bT^{a_1}_{i_1}$) up until the operator $\bT^{a_\ell}_{i_\ell}$. The operator $(\bT^{a_\ell}_{j_\ell})^{\dag}$ is then immediately inserted after $\bT^{a_\ell}_{i_\ell}$. That is to say as soon as the virtual gluon $k_\ell$ is emitted (by leg $k_{i_\ell}$) it gets absorbed (by leg $k_{j_\ell}$). The ordering of color operators then continues as usual. Schematically we write:
\begin{align}
\cC^{i_1 \hdots i_\ell \hdots i_m}_{j_1 \hdots j_\ell \hdots j_m}=&\,\frac{(-1)^{m-1}}{N_c} \bra{c_a',c_a'} \left(\bT^{a_1}_{j_1}\right)^\dag \cdots \left(\bT^{a_m}_{j_m}\right)^\dag\cdot \notag \\
&\cdot \bT^{a_m}_{i_m}\cdots \left(\bT^{a_\ell}_{j_\ell}\right)^\dag \bT^{a_\ell}_{i_\ell} \cdots \bT^{a_1}_{i_1} \ket{c_a,c_a},
\label{eq:EikAmp_VirTreat1}
\end{align}
where the ``missing'' factor $(-1)$, relative to Eq. \eqref{eq:EikColorFactor}, is due to the fact that virtual gluons have no polarization vector. Notice also that, in going from Eq. \eqref{eq:EikColorFactor} to Eq. \eqref{eq:EikAmp_VirTreat1}, the color operator $(\bT^{a_\ell}_{j_\ell})^\dag$ has been moved from its position in the conjugated amplitude (bra) to the unconjugated one (ket). Since the virtual gluon $k_\ell$ can also be attached in the conjugated amplitude, the opposite scenario must also be taken into account. That is, we also add the case in which the operator $\bT^{a_\ell}_{i_\ell}$ is placed next to $(\bT^{a_\ell}_{j_\ell})^\dag$ in the conjugated amplitude:\footnote{For the case of a single virtual gluon, this results in an overall identical contribution to that in Eq. \eqref{eq:EikAmp_VirTreat1}. Thus one can consider only the first scenario and multiply by a factor 2.}
\begin{align}
\widetilde{\cC}^{i_1 \hdots i_\ell \hdots i_m}_{j_1 \hdots j_\ell \hdots j_m} =&\, \frac{(-1)^{m-1}}{N_c} \bra{c_a',c_a'} \left(\bT^{a_1}_{j_1}\right)^\dag \cdots \left(\bT^{a_\ell}_{j_\ell}\right)^\dag \cdot \notag \\
& \cdot \bT^{a_\ell}_{i_\ell}\cdots \left(\bT^{a_m}_{j_m}\right)^\dag \bT^{a_m}_{i_m} \cdots \bT^{a_1}_{i_1} \ket{c_a,c_a}.
\label{eq:EikAmp_VirTreat2}
\end{align}

The attachment of the virtual gluon $k_\ell$ must be spanned over all possible harder real parton pairs (dipoles $(i_\ell j_\ell)$) in the event without double counting (i.e., we set $i_\ell>j_\ell$). The virtual gluon $k_\ell$ itself does not, however, play the role of an emitter for softer gluons and must thus be excluded from the list of emitters $U_{n-1}$ in Eq. \eqref{eq:EikAmpSq_Final}.

For more than one virtual gluon, one proceeds in an analogous manner to that presented above (for one virtual gluon). For each virtual gluon one places the color operators associated with its emitting dipole next to each other in the ordered expression of the color factor \eqref{eq:EikColorFactor} (one time in the bra-side and another in the ket-side) and removes a relative factor of $(-1)$. Additionally, virtual gluons cannot be emitters and are consequently omitted from the list of emitters $U_m$.

Based on this observation, we shall show in Sec. \ref{sec:Examples} that only the ``all-gluons-real'' ($X = \mr{RR \hdots R}$) squared amplitude needs to be explicitly evaluated. The squared amplitudes for all other configurations may be deduced from the latter without any further rerun of the {\tt EikAmp} program. This is straightforwardly achieved given the analytical compact forms of the eikonal squared amplitudes that we shall present therein.

\subsection{The EikAmp program}

The eikonal amplitude squared given in Eq. \eqref{eq:EikAmpSq_Final} is computed using a {\it Mathematica} program ``{\tt EikAmp}'', which we have developed for this very purpose. The program starts by calculating the color factor \eqref{eq:EikColorFactor} using the {\tt ColorMath} package. The main idea of the program is to use the ``{\tt For loop}'' to account for all possible dipoles. The outer-most loop corresponds to the hardest gluon and picks up the first leg of the dipole $(ab)$. The next-to-outer-most loop runs over the legs $\{a, b, 1\}$ picking up one at each run. The inner loops work in the same way.

Once the product of the color matrices for the amplitude is determined, the program moves onto computing the corresponding product of color matrices for the conjugate amplitude. Since each gluon connects the legs of a given dipole then in this second step the program basically picks up the second leg to form the emitting dipoles. The number of possible dipoles that can emit the $m^{\mr{th}}$ gluon is $m(m+1)/2$ (assuming that $(ij)$ and $(ji)$ represent the same  dipole). After that, the program calls the {\tt ColorMath} function {\tt CSimplify} to carry out the trace of the product of the color matrices. The final result is a scalar written in terms of $N_c$, which is then decomposed into a product of $\CF$ and $\CA$ factors. Moreover, the program associates with every dipole $(ij)$ emitting gluon $k$ the antenna function $w^k_{ij}$. Once the program has ended one may collect terms with a common color factor.

The above procedure may be repeated for each possible gluon configuration $X$, taking into account the symmetries mentioned above to avoid unnecessary repetitions. Full details, including the {\tt c++} and/or {\tt fortran} versions of the program, are deferred to future publications.

The output squared amplitude of the {\tt EikAmp} program at a given loop order is a complicated sum of all possible Feynman-diagram contributions, which renders the result impractical to some extent. Based on the pattern and symmetry observed at two loops, as we shall see below in Sec. \ref{sec:Examples}, it is possible to write the output amplitudes squared at higher-loop orders in a \emph{compact} form that is of more practical usage. For example, it allows us to deduce squared amplitudes for configurations involving virtual gluons from the all-gluons-real amplitude squared without any explicit calculations. Furthermore, it enables us to separate the finite-$N_c$ corrections from the large-$N_c$ contributions, thus paving the way for later analyzes of the two contributions.

\subsection{Large-$\bm N_c$ limit}

In the large-$N_c$ limit all Feynman diagrams that have a non-planar topology are discarded as their corresponding cross-sections are suppressed by $1/N_c^2$ \cite{Hooft1974461}.\footnote{Non-planar diagrams are those with crossing color lines.} Consequently, both the number of Feynman diagrams, to be evaluated at a given order, and the color structure simplify considerably. In fact, the color factor \eqref{eq:EikColorFactor} associated with the emission of $m$ real gluons reduces to:
\begin{align}
\cC^{i_1 \hdots i_m}_{j_1 \hdots j_m} = \left(\frac{N_c}{2}\right)^m,
\label{eq:EikColorFactor_LargeNc}
\end{align}
for all possible values of the indices $i_1, \hdots, i_m, j_1, \hdots, j_m$ (except for non-planar diagrams in which case the color factor is zero). The factor of $1/2$ in the above expression may straightforwardly be deduced recalling that the color factor for the first emission, in the large-$N_c$ limit, is $\CF = N_c/2$. All subsequent emissions have an $N_c/2$ factor.

For the kinematical part of the eikonal amplitude squared one should note that, since the color factor at large $N_c$ is a simple scalar given in Eq.  \eqref{eq:EikColorFactor_LargeNc}, one may safely set $i_n > j_n$ in Eq. \eqref{eq:EikAmpSq_Final} and multiply the whole expression by $2^m$. This factor then cancels out the $(1/2)^m$ in Eq. \eqref{eq:EikColorFactor_LargeNc}. Hence, one can pull  $N_c^m$ out of the definition of the eikonal amplitude squared $\cW^{\R\R\hdots\R}_{12\hdots m}$ in Eq. \eqref{eq:EikAmpSq_Final} and absorb it into the coupling such that $\asb = N_c\, \as/\pi$. The remaining, purely angular, part of the eikonal amplitude squared \eqref{eq:EikAmpSq_Final}, and after discarding non-planar configurations, should reduce to the well-known form \cite{Bassetto:1984ik, Fiorani:1988by, Banfi:2002hw}:\footnote{This formula is only valid for the all-gluons-real configuration, and to obtain the corresponding amplitudes squared for configurations with virtual gluons one should follow the steps outlined in Ref. \cite{Schwartz:2014wha}.}
\begin{align}
\widetilde{\cW}^{\R\R\hdots \R}_{12\hdots m} = \left(\prod_{i=1}^m k_{ti}^2\right) \sum_{\pi_m} \frac{ \left(p_a \cdot p_b\right) }{ (p_a \cdot k_1) (k_1 \cdot k_2) \cdots (k_m \cdot p_b)}\,,
\label{eq:EikAmpSq_LargeNc_A}
\end{align}
where the sum is over all $m!$ permutations of the set of momenta $\{k_1, k_2,\hdots, k_m \}$. Eq. \eqref{eq:EikAmpSq_LargeNc_A} has been explicitly verified up to five loops using the {\tt EikAmp} program via the simple substitutions $\CF \to N_c/2$ and $\CA \to N_c$.

In the next section we report on our results for the eikonal amplitudes squared at finite $N_c$ up to five loops for all possible gluon configurations $X$.

\section{Examples}
\label{sec:Examples}

\subsection{One and two loops}
\label{subsec:OneTwoLoops}

We start by presenting the already-known results at one- and two-loop orders. Before doing so we define the following generalized $n$-loop antenna functions which will prove useful in what follows below:
\begin{subequations}
\begin{align}
\cA_{ab}^{ij} &= w^i_{ab} \left(w^j_{ai} + w^j_{ib} - w^j_{ab} \right), \\
\cB^{ijk}_{ab} &= w^i_{ab} \left(\cA^{jk}_{ai} + \cA^{jk}_{ib} - \cA^{jk}_{ab} \right), \\
\cC^{ijk\ell}_{ab} &= w^i_{ab} \left(\cB^{jk\ell}_{ai} + \cB^{jk\ell}_{ib} - \cB^{jk\ell}_{ab} \right), \\
\cD^{ijk\ell m}_{ab} &= w^i_{ab} \left(\cC^{jk\ell m}_{ai} + \cC^{jk\ell m}_{ib} - \cC^{jk\ell m}_{ab} \right).
\end{align}
\label{eq:Antennas_defs}
\end{subequations}
First, it should be understood that all sub and superscript indices are always different from each other. In fact, superscript indices are ordered according to the energy ordering of the corresponding gluons, as it should be clear by expanding higher-loop antenna functions in terms of the two-loop antenna $\cA_{ij}^{k\ell}$. Secondly, whenever we write any one of the above antennae with an overhead bar, such as $\cAb_{ab}^{ij}$, then we mean the same definition but with the $w^i_{ab}$ divided out. That is, $\cAb^{ij}_{ab} = \cA^{ij}_{ab}/w^i_{ab}$, $\cBb^{ijk}_{ab} = \cB^{ijk}_{ab}/w^i_{ab}$, and so on.

The one-loop eikonal amplitudes squared are quite simple and read:
\begin{subequations}\label{eq:EikAmp_1Loop}
\begin{align}
\cW^\R_{1} &= \CF \left(w^1_{ab} + w^1_{ba} \right) =  2\,\CF\, w^1_{ab}\,, \\
\cW^\V_1 &= - \cW^\R_1\,.
\end{align}
\end{subequations}
They represent an independent emission of gluon $k_1$ by the dipole $(ab)$. Notice that, since the dipoles $(ab)$ and $(ba)$ are associated with the same color factor, then we fixed the ordering of the legs, picking up $(ab)$ for the one-loop antenna $w^1_{ab}$, and multiplied by a factor of $2$. Moreover, since each $n$-loop antenna function \eqref{eq:Antennas_defs} involves products of $n$ one-loop antennae, and if each of which is symmetric under the interchange of its legs ($w^k_{ij} = w^k_{ji}$),\footnote{We shall see, however, that the latter condition does not hold starting at four loops.} then one may fix the ordering of the legs for each dipole such that the $n$-loop amplitude squared becomes proportional to the factor $2^n$. For instance, the two-loop amplitude squared containing  $\cA^{ij}_{ab}$ is proportional to $2^2$, the three-loop amplitude squared containing the antenna $\cB^{ijk}_{ab}$ to $2^3$, etc.

\begin{figure}[ht]
\centering
\includegraphics[width=0.3\textwidth]{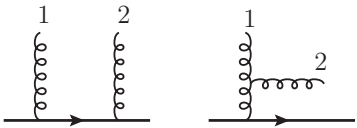}
\caption{A diagrammatic representation of the reducible (left) and irreducible (right) contributions to the two-loops eikonal amplitude squared $\cW^{\R\R}_{12}$.\label{fig:EikAmp_2Loop}}
\end{figure}

The two-loop squared amplitudes are given by:
\begin{subequations}
\begin{align}
\cW^{\R\R}_{12} &= \cW^\R_1 \cW^\R_2 + \cWI^{\R\R}_{12}\,,\quad & \cW^{\R\V}_{12} = - \cW^{\R\R}_{12}\,, \\
\cW^{\V\R}_{12} &=- \cW^{\R}_1 \cW^\R_2\,, &  \cW^{\V\V}_{12} = - \cW^{\V\R}_{12}\,.
\end{align}\label{eq:EikAmp_2Loop}
\end{subequations}
Focusing on the all-gluons-real amplitude squared $\cW^{\R\R}_{12}$, the first \emph{reducible} contribution $\cW^\R_1 \cW^\R_2$ is nothing but successive independent emissions of gluons $k_1$ and $k_2$. The second contribution $\cWI^{\R\R}_{12}$ is \emph{irreducible} in the sense that it is not related to the one-loop amplitude squared. It reads:
\begin{align}
\left[\frac{1}{2^2}\right] \cWI^{\R\R}_{12} = \frac{1}{2}\,\CF\CA\, \cA^{12}_{ab} \,,
\label{eq:EikAmp_2Loop_Irred}
\end{align}
where the origin of the factor $2^2$ has been discussed above. It has been moved to the left-hand side of Eq. \eqref{eq:EikAmp_2Loop_Irred} so as to emphasize the color structure of a given individual irreducible diagram. Fig. \ref{fig:EikAmp_2Loop}
is a diagrammatic representation of the two-loops eikonal amplitude squared. The antenna function $\cA^{12}_{ab}$ represents a two-parton \emph{cascade} emission. Unlike the aforementioned reducible contribution, which has a total of two angular singularities corresponding to gluons $k_1$ and $k_2$ being collinear to either $q$ and/or $\qbar$, the antenna function $\cA_{ab}^{12}$ has only one angular singularity corresponding to gluons $k_1$ and $k_2$ being on top of each other. For particular phase-space configurations this antenna function may be integrated out producing no extra logarithms in the cross-section (see for instance Ref. \cite{Khelifa-Kerfa:2015mma}), and contributing solely single energy logarithms (originating from gluons $k_1$ and $k_2$ being soft). This means that it yields the single-logarithmic contribution, $(\as\,L)^2$, to the integrated cross-section of an exclusive observable (with the reducible term $\cW^\R_1 \cW^\R_2$ resulting in up to double-logarithmic contributions $(\as\, L^2)^2$). It is worthwhile to mention that the two-loop antenna $\cA_{ij}^{k\ell}$ is different from the ``remainder'' function considered in Ref. \cite{Dokshitzer:1991wu}, in the sense that the latter has no singular dependence on angles (a property referred to as \emph{friability}) as well as being integrable over the directions of all gluons involved (a property dubbed \emph{ideality} in \cite{Dokshitzer:1991wu}).

\subsection{Three loops}

There are $2^3$ squared amplitudes to be evaluated at this order which reduces by symmetry to only $4$. These are given by:
\begin{subequations}
\begin{align}
\cW^{\R\R\R}_{123} &=   \prod_{i=1}^3 \cW^{\R}_i + \sum_{ijk=1}^3 \cW^\R_i\,\cWI^{\R\R}_{jk} + \cWI^{\R\R\R}_{123}\,,\label{eq:EikAmp_3Loopa}\\
\cW^{\R\V\R}_{123} &= - \prod_{i=1}^3 \cW^{\R}_i - \sum_{ik=2}^3 \cW^\R_i\,\cWI^{\R\R}_{1k} + \cWI^{\R\V\R}_{123}\,,\label{eq:EikAmp_3Loopb}\\
\cW^{\V\R\R}_{123} &= - \prod_{i=1}^3 \cW^{\R}_i - \cW^\R_1\,\cWI^{\R\R}_{23}\,,\label{eq:EikAmp_3Loopc} \\
\cW^{\V\V\R}_{123} &=   \prod_{i=1}^3 \cW^{\R}_i \,,
\end{align}
\label{eq:EikAmp_3Loop}
\end{subequations}
and as stated before:
\begin{equation}
\cW^{\mr{x}_1\mr{x}_2\V} = - \cW^{\mr{x}_1\mr{x}_2\R}\,.
\end{equation}
The three-loops irreducible terms read:
\begin{subequations}
\begin{align}
\left[\frac{1}{2^3}\right] \cWI^{\R\R\R}_{123} &=   \frac{1}{4}\,\CF\CAsq \left(\cA^{12}_{ab} \cAb^{13}_{ab} + \cB_{ab}^{123} \right), \label{eq:EikAmp_3Loop_Irred0}\\
\left[\frac{1}{2^3}\right] \cWI^{\R\V\R}_{123} &= - \frac{1}{4}\,\CF\CAsq\, \cA^{12}_{ab} \cAb^{13}_{ab}\,.\label{eq:EikAmp_3Loop_Irred}
\end{align}\label{eq:EikAmp_3Loop_Irred00}
\end{subequations}
We remind the reader that, similar to the antenna functions, throughout this paper whenever we show a sum over several indices of amplitudes squared, such as those in Eqs. \eqref{eq:EikAmp_3Loopa} and \eqref{eq:EikAmp_3Loopb}, then these indices may only take different values at once. Furthermore, and to avoid double counting, whenever we write $\cW^{\X}_{ij}$ (or $\cWI^{\X}_{ij}$) it is implied that $i<j$, which reflects the energy ordering of the corresponding gluons $\omega_i > \omega_j$. Likewise for three- and higher-loop amplitudes squared, the indices are always ordered ascendantly from left to right.

\begin{figure}[ht]
\centering
\includegraphics[width=.48\textwidth]{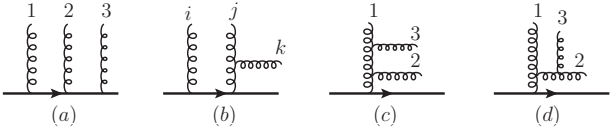}
\caption{A diagrammatic representation of the reducible and irreducible contributions to the three-loops eikonal amplitude squared $\cW^{\R\R\R}_{123}$.\label{fig:EikAmp_3Loop}}
\end{figure}

The various contributions to the all-gluons-real amplitude squared $\cW^{\R\R\R}_{123}$ are depicted in Fig. \ref{fig:EikAmp_3Loop}.
The first term, diagram $(a)$, represents as usual the successive independent emission of three gluons. The second, diagram $(b)$, is an \emph{interference} between the one-loop amplitude squared $\cW^\R_i$ and the two-loops irreducible term $\cWI^{\R\R}_{jk}$. These two terms together form the reducible contribution at three loops, which may be recast as follows:
\begin{align}
\cWR^{\R\R\R}_{123} &= \frac{1}{s_3} \sum_{ijk = 1}^3 \cW^\R_i \, \cW^{\R\R}_{jk} \notag \\
&= \prod_{i=1}^3 \cW^\R_i + \sum_{ijk=1}^3 \cW^\R_i \cWI^{\R\R}_{jk} \,,\label{eq:EikAmp_3Loop_Red}
\end{align}
where the coefficient $1/s_3$ is there to cancel out any repeated occurrences of the same contribution such that after expanding the sum one recovers the first two terms of Eq. \eqref{eq:EikAmp_3Loopa}. With the above definition, i.e., Eq. \eqref{eq:EikAmp_3Loop_Red}, the all-gluons-real amplitude squared in Eq. \eqref{eq:EikAmp_3Loopa} may be written in the compact form:
\begin{align}
\cW^{\R\R\R}_{123} = \cWR^{\R\R\R}_{123} + \cWI^{\R\R\R}_{123}\,. \label{eq:EikAmp_3loop_BlockForm}
\end{align}
In fact, and as we have already seen at two loops \eqref{eq:EikAmp_2Loop} and shall see at four  and five loops, one anticipates that all eikonal amplitudes squared may be represented in an analogous form to that of Eq. \eqref{eq:EikAmp_3loop_BlockForm}.

Turning to the three-loops irreducible contribution $\cWI^{\R\R\R}_{123}$ in Eq. \eqref{eq:EikAmp_3Loop_Irred0}, we observe that it consists of two terms, illustrated respectively by diagrams $(c)$ and $(d)$ of Fig. \ref{fig:EikAmp_3Loop}. The first term $\cA^{12}_{ab} \cAb^{13}_{ab}$ may be envisaged as a product of two two-loops antenna functions with the second antenna not connected to the quark line as shown below in Fig. \ref{fig:EikAmp_3Loop1}.
\begin{figure}[ht]
\centering
\includegraphics[width=.48\textwidth]{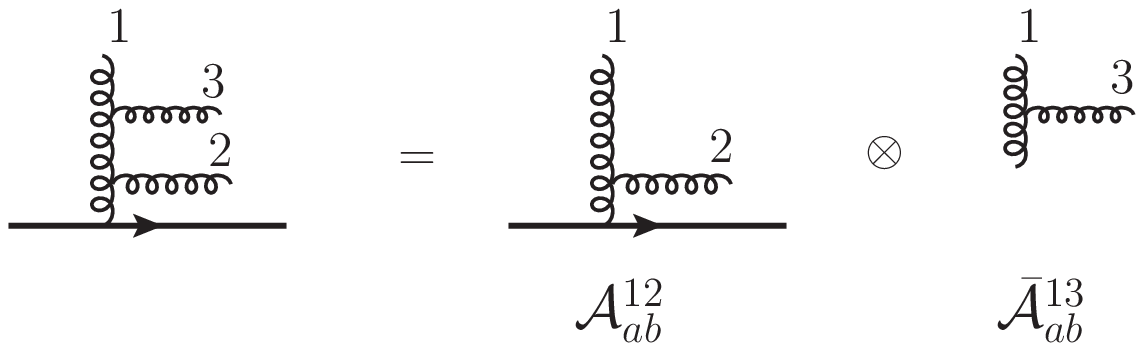}
\caption{The term $\cA^{12}_{ab} \cAb^{13}_{ab}$ of the irreducible contribution \eqref{eq:EikAmp_3Loop_Irred0}.\label{fig:EikAmp_3Loop1}}
\end{figure}
One can equally think of it as $\cAb^{12}_{ab} \cA^{13}_{ab}$, i.e., the first antenna --- including gluon $k_2$ --- is not connected to the quark line whilst the second antenna $\cA^{13}_{ab}$ is. Furthermore, this term may also be thought of as successive independent emissions of gluons $k_2$ and $k_3$ off the hardest gluon $k_1$, which in turn was radiated off by the $(ab)$ dipole. The second term $\cB^{123}_{ab}$ represents a three-parton cascade emission, and is non-vanishing only for the all-gluons-real amplitude squared. The contributions $(c)$ and $(d)$ are neither friable nor ideal as they contain, in addition to the ``remainder'' of Ref. \cite{Dokshitzer:1991wu}, terms that are singular in angles separating the softest gluon $k_3$ and the harder gluons $k_1$ and $k_2$.

In fact, the contribution $\cWI^{\R\R\R}_{123}$ is \emph{not} completely irreducible since it consists of terms resembling those already seen at two loops. Recall that the cascade term $\cB^{123}_{ab}$ is written in terms of the two-loops antenna $\cA^{ij}_{ab}$ (see Eqs. \eqref{eq:Antennas_defs}). It is due to this ``deluding'' irreducibility that at this order large-$N_c$ calculations yield identical results to those shown in Eq. \eqref{eq:EikAmp_3Loop}. In other words, there are no finite-$N_c$ corrections up to this order. The first \emph{genuine} irreducible contributions pop up at four loops as we show below. Moreover, it is the first order where finite-$N_c$ corrections do appear.

\subsection{Four loops}

There are $2^4$ squared amplitudes to be evaluated at this order which reduces, due to symmetry, to only $8$.\footnote{It is mentioned in Ref. \cite{Dokshitzer:1991wu} that the four-loop squared amplitude was computed in Ref. \cite{Dokshitzer:1984xx} using the ``probabilistic scheme'' developed in the former reference. We have not been able to obtain a copy of the latter reference \cite{Dokshitzer:1984xx} to compare it with our results.}  In fact, it is sufficient to only explicitly present and discuss the features of the all-gluons-real amplitude squared as it contains all possible contributions at a given order. The rest of the squared amplitudes can be deduced quite straightforwardly as we shall mention towards the end of this subsection.

\begin{figure}[ht]
\centering
\includegraphics[width=0.48\textwidth]{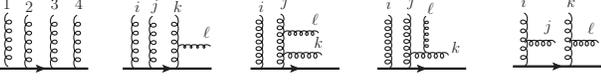}
\caption{A diagrammatic representation of the reducible contributions to the four-loops eikonal amplitude squared.\label{fig:EikAmp_4Loop_red}}
\end{figure}

The all-gluons-real eikonal amplitude squared at four loops may be written as:
\begin{align}
\cW^{\R\R\R\R}_{1234} = \cWR^{\R\R\R\R}_{1234} + \cWI^{\R\R\R\R}_{1234}\,,\label{eq:EikAmp_4Loop}
\end{align}
where the reducible contribution is given by:
\begin{align}
\cWR^{\R\R\R\R}_{1234} =&\, \frac{1}{s_4} \sum_{ijk\ell=1}^4 \left[\cW^\R_i\, \cW^{\R\R\R}_{jk\ell} + \cW^{\R\R}_{ij}\, \cW^{\R\R}_{k\ell} \right]\notag \\
=&\, \prod_{i=1}^4 \cW^\R_i + \sum_{ijk\ell=1}^4 \left[\cW^\R_i\, \cW^\R_j \,\cWI^{\R\R}_{k\ell} \,\Theta^{i<j} \right.+\notag\\
&\,\left.+  \cW^\R_i \,\cWI^{\R\R\R}_{jk\ell} + \cWI^{\R\R}_{ij}\, \cWI^{\R\R}_{k\ell} \, \Theta^{i<k}  \right], \label{eq:EikAmp_4Loop_Red}
\end{align}
where in analogy to three loops, the factor $1/s_4$ divides out any repeated counting of the same contribution. The step functions $\Theta^{x<y} \equiv \Theta(y-x)$ that appear in the above expression are there to avoid double counting of the terms. The various terms in Eq. \eqref{eq:EikAmp_4Loop_Red} are shown schematically in Fig. \ref{fig:EikAmp_4Loop_red}.

\begin{figure}[ht]
\centering
\includegraphics[width=0.48\textwidth]{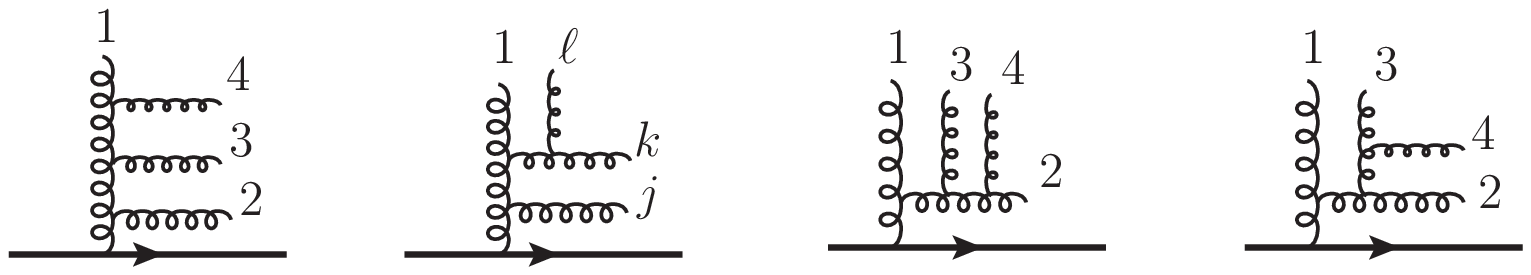}
\caption{A diagrammatic representation of the irreducible contributions \eqref{eq:EikAmp_4Loop_Irred_Red} to the four-loops eikonal amplitude squared.\label{fig:EikAmp_4Loop_irred}}
\end{figure}

The irreducible contribution may be split into two parts:
\begin{align}
\cWI^{\R\R\R\R}_{1234} = \cWIR^{\R\R\R\R}_{1234} + \cWII^{\R\R\R\R}_{1234} \,,  \label{eq:EikAmp_4Loop_Irred}
\end{align}
where the first part though irreducible resembles patterns seen at previous orders. It reads:
\begin{multline}
\left[\frac{1}{2^4}\right]\cWIR^{\R\R\R\R}_{1234} =  \frac{\CF\CAcub}{8}\Bigg(\cA^{12}_{ab} \cAb^{13}_{ab} \cAb^{14}_{ab}  + \sum_{jk\ell=2}^4 \cA^{1j}_{ab} \cBb^{1k\ell}_{ab} + \\
+\fA_{ab}^{1234}  +  \cC^{1234}_{ab} \Bigg)\,, \label{eq:EikAmp_4Loop_Irred_Red}
\end{multline}
with:
\begin{align}\label{eq:Pseudo-Antennas_Defs0}
\fA^{ijk\ell}_{ab} = w^i_{ab} \left( \cA^{jk}_{ai} \cAb^{j\ell}_{ai} + \cA^{jk}_{ib} \cAb^{j\ell}_{ib} - \cA^{jk}_{ab} \cAb^{j\ell}_{ab} \right).
\end{align}
The various terms in the expression \eqref{eq:EikAmp_4Loop_Irred_Red} are shown, respectively, from left to right below (Fig. \ref{fig:EikAmp_4Loop_irred}).

The second part is the new genuinely-irreducible contribution that contains the first finite-$N_c$ correction:
\begin{align}
\cWII^{\R\R\R\R}_{1234} = 4\, \CF\CAsq \left(\CF - \frac{\CA}{2} \right) \cN^{\R\R\R\R}_{1234}\,, \label{eq:EikAmp_4Loop_Irred_Irred}
\end{align}
where $\cN^{\X}_{1234}$ is the angular function of the irreducible finite-$N_c$ contribution. At four loops it is given by:
\begin{align}
\cN^{\R\R\R\R}_{1234} = \bA_{ab}^{1234} \,, \label{eq:EikAmp_4Loop_Irred_Irred2}
\end{align}
where we have defined the following ``pseudo-antenna'' function:
\begin{align}
\bA_{ab}^{ijk\ell} =w^i_{ab} &\left[ w^j_{ai}\, T^{bj}_{ai}(k) U^{aj}_{bi}(\ell) + w^j_{ib}\, T^{aj}_{bi}(k) U^{bj}_{ai}(\ell)-  \right.\notag\\
&\left.\,\,- w^j_{ab}\, T^{ij}_{ab}(k) U^{ij}_{ab}(\ell)  \right] + k \leftrightarrow \ell \,,
\label{eq:Pseudo-Antennas_Defs}
\end{align}
which in turn is written in terms of the $t$- and $u$-channel functions:
\begin{subequations}
\begin{align}
T_{ij}^{k\ell}(n) &= w^n_{ij} + w^n_{k\ell} - w^n_{ik} - w^n_{j\ell} = \cAb_{ik}^{\ell n} - \cAb_{ij}^{\ell n} \,,\\
U_{ij}^{k\ell}(n) &= w^n_{ij} + w^n_{k\ell} - w^n_{i\ell} - w^n_{jk} = \cAb_{jk}^{\ell n} - \cAb_{ij}^{\ell n}\,.
\end{align}
\label{eq:T-U_Channel_Funs}
\end{subequations}
It is interesting to note that analogous functions to the above $t$- and $u$-channel terms were derived in Ref. \cite{Dokshitzer:2005ek} while considering the emission of a soft gluon off an ensemble of four colored hard partons --- in hadron-hadron collisions. They were proven therein to be integrable over the direction of the emitted softest gluon (represented by the superscript index $n$ in the above).

Furthermore, it is worth mentioning that, as stated previously, the finite-$N_c$ correction \eqref{eq:EikAmp_4Loop_Irred_Irred} has the peculiar feature that it is \emph{not} symmetric under the permutation of the gluon indices except for two: ($1 \leftrightarrow 2$) and ($3 \leftrightarrow 4$). For instance, one may easily verify that:
\begin{align}
\cN^{\R\R\R\R}_{1234} \neq  \cN^{\R\R\R\R}_{1324} \neq  \cN^{\R\R\R\R}_{4123}\,.
\label{eq:EikAmp_FinteNc_NonSym}
\end{align}
This is in contrast to the large-$N_c$ contributions \eqref{eq:EikAmp_4Loop_Red} and \eqref{eq:EikAmp_4Loop_Irred_Red}, which are totally symmetric under any given permutation.

In addition to the breaking of the ``Bose symmetry'' mentioned above, corresponding to permutations of emitted gluons, there is another broken symmetry at four loops that we refer to as ``mirror symmetry''. It may be summarized as follows: there are (eikonal) Feynman diagrams which correspond to color factors that are \emph{not symmetric} under the interchange of the emitters of one or more gluons (or equivalently, interchange of the legs of one or more emitting dipoles). It turns out that only diagrams which contribute to the finite-$N_c$ terms $\cN_{1234}^X$ break this symmetry. The squared amplitudes that break the mirror symmetry at four loops are $\cW^{\R\R\R\R}_{1234}$ and $\cW_{1234}^{\R\R\V\R}$, i.e., those with gluons $k_1$ and $k_2$ real. The remaining squared amplitudes, such as $\cW^{\R\V\R\R}_{1234}$ and $\cW^{\R\V\V\R}_{1234}$, are symmetric under both mirror and Bose symmetries.

While the mirror-symmetry breaking at five loops results in \emph{fractional} coefficients multiplying color factors, as we shall see in the next subsection, the four-loops mirror-symmetry breaking does not. The reason for this could be that the permutation of dipole legs in non-planar Feynman diagrams at four loops yields either an identical color factor or zero.\footnote{We have tested this with a number of Feynman diagrams (at four loops).} Thus adding up all diagram contributions results in an overall \emph{integer} coefficient multiplying the color factor (associated with the said diagram).

Notice also that since the large- and finite-$N_c$ contributions to the eikonal amplitude squared are separated from each other one may accordingly write:
\begin{align}
\cW^{\R\R\R\R}_{1234} = \cW^{\R\R\R\R,\, \mr{lN_c}}_{1234} + \cW^{\R\R\R\R,\, \mr{fN_c}}_{1234}\,.
\label{eq:EikAmp_4loop_Lnc-Fnc_Facorised}
\end{align}

\subsubsection{Amplitudes squared for other gluon configurations}
\label{subsec:VirtualAmps}

In order to compute the remaining seven squared amplitudes (and by symmetry the other seven corresponding to the softest gluon being virtual) it is sufficient to set to zero the antenna functions \eqref{eq:Antennas_defs}, \eqref{eq:Pseudo-Antennas_Defs} and \eqref{eq:T-U_Channel_Funs} for which the virtual gluons play the role of emitters, in the expression of the all-gluons-real squared amplitude \eqref{eq:EikAmp_4Loop}. For instance, to find the form of the eikonal amplitude squared $\cW^{\R\R\V\R}_{1234}$, with gluon $k_3$ being virtual, we simply write:
\begin{align}
\cW^{\R\R\V\R}_{1234} = -\left.\cW^{\R\R\R\R}_{1234}\right|_{S_3 \to 0} \,, \qquad \cW^{\R\R\V\V}_{1234} = - \cW^{\R\R\V\R}_{1234}\,,
\end{align}
where $S_\alpha$ is the set of all possible antennae for which the virtual gluon $k_\alpha$ acts as an emitter for softer gluons:
\begin{align}
S_\alpha =\Big\{&  \cA_{\alpha j}^{k\ell}, \cA_{i\alpha}^{k\ell}, \cA_{ij}^{\alpha\ell}, \cB_{\alpha j}^{k\ell m},\cB_{i\alpha}^{k\ell m},\cB_{ij}^{\alpha\ell m},\cB_{ij}^{k \alpha m}, \cC_{\alpha j}^{k\ell m n}, \notag\\
& \cC_{i\alpha}^{k\ell m n}, \cC_{ij}^{\alpha\ell m n}, \cC_{ij}^{k \alpha m n}, \cC_{ij}^{k\ell \alpha n},T_{\alpha j}^{k\ell}(m), T_{i\alpha}^{k\ell}(m),  \notag\\
&  T_{ij}^{\alpha\ell}(m), T_{ij}^{k \alpha}(m),U_{\alpha j}^{k\ell}(m), U_{i\alpha}^{k\ell}(m), U_{ij}^{\alpha\ell}(m), \notag\\
& U_{ij}^{k \alpha}(m), \fA_{\alpha j}^{k\ell m n}, \fA_{i\alpha}^{k\ell m n}, \fA_{ij}^{\alpha\ell m n}, \fA_{ij}^{k \alpha m n}, \bA_{\alpha j}^{k\ell m n}, \notag\\
& \bA_{i\alpha}^{k\ell m n}, \bA_{ij}^{\alpha\ell m n}, \bA_{ij}^{k \alpha m n} \Big\}\,. \label{eq:VirSub_4Loop_Va}
\end{align}
The statement $S_\alpha \to 0$ means that each element in the set $S_\alpha$ is set to zero.

Remark that within the {\tt EikAmp} program it is not, however, an easy task to set $S_\alpha \to 0$ since one would have to write the cumbersome output in terms of the various antenna functions (see for instance Eqs. \eqref{eq:Antennas_defs} and \eqref{eq:T-U_Channel_Funs}) before setting any of them to zero. Whether this is doable will be investigated in our forthcoming {\tt EikAmp} paper. It should be noted meanwhile that it is not sufficient to simply set the dipole antenna $w^k_{ij} \to 0$ if either gluon $i$ or $j$ is virtual in order to find the corresponding amplitude squared.

Performing the substitution $S_3 \to 0$ one finds:
\begin{align}
\cW^{\R\R\V\R}_{1234} = \cWR^{\R\R\V\R}_{1234} + \cWI^{\R\R\V\R}_{1234}\,,
\end{align}
where for the reducible contribution:
\begin{align}
\cWR^{\R\R\V\R}_{1234} =\,& - \prod_{i=1}^4 \cW^\R_i - \sum_{ij,(k\neq 3),\ell=1}^4 \left[\cW^\R_i \, \cW^\R_j \, \cWI^{\R\R}_{k\ell}\, \Theta^{i<j} +\right.\notag\\
 &+  \cW^\R_i\, \cWI^{\R\R\R}_{jk\ell} - \cW^\R_2\, \cWI^{\R\V\R}_{134} - \cW^\R_1\, \cWI^{\R\V\R}_{234} + \notag\\
 &\left.+ \cWI^{\R\R}_{ij}\, \cWI^{\R\R}_{k\ell} \,\Theta^{i<k} \right],
\end{align}
where we note that $\cWI^{\R\V\R}_{134}$ and $\cWI^{\R\V\R}_{234}$ already have a relative minus sign (see Eq. \eqref{eq:EikAmp_3Loop_Irred}). The irreducible contribution is:
\begin{align}
\cWI^{\R\R\V\R}_{1234}  =  \cWIR^{\R\R\V\R}_{1234} + \cWII^{\R\R\V\R}_{1234}\,,
\end{align}
with:
\begin{subequations}
\begin{align}
\left[\frac{1}{2^4} \right] \cWIR^{\R\R\V\R}_{1234} &= - \frac{\CF\CAcub}{8} \bigg[\cA^{12}_{ab} \cAb^{13}_{ab} \cAb^{14}_{ab} +\fA_{ab}^{1234}+ \notag\\
&\qquad\qquad\quad\,\, +\sum_{j\ell=3}^4 \cA^{1j}_{ab} \cBb^{12\ell}_{ab} \bigg]\,, \\
\cWII^{\R\R\V\R}_{1234} &= - \cWII^{\R\R\R\R}_{1234}\,.
\end{align}
\end{subequations}
All other amplitudes squared may be obtained in a similar fashion. One simply replaces $\alpha$ in \eqref{eq:VirSub_4Loop_Va} by the appropriate virtual-gluon number. For more than one virtual gluon, the elements of all sets $S_\beta$ corresponding to all virtual gluons $k_{\beta}$, with $\beta = 1,2, \hdots,$ must be set to zero simultaneously.

Moreover, one may also observe the following alternative reduced expressions relating virtual amplitudes squared at four loops to the amplitude squared at two and three loops:
\begin{subequations}
\begin{align}
\cW_{1234}^{\R\V\mr{x_3}\R} & = \frac{\cW^{\R\V}_{12}}{\cW^{\R}_1}\,\cW_{134}^{\R\mr{x_3}\R}\,,  \\
\cW_{1234}^{\V\mr{x_2}\mr{x_3}\R} & = \cW^{\V}_{1}\,\cW_{234}^{\mr{x_2}\mr{x_3}\R}\,.
\end{align}
\end{subequations}

In the next subsection we present and discuss the expression of the five-loops eikonal amplitude squared.

\subsection{Five loops}
\label{sebsec:FiveLoops}

At five-loops order there is a total of $2^5$ squared amplitudes to be determined, half of which are deduced by symmetry. Out of these only the all-gluons-real squared amplitude needs to be actually calculated. The other squared amplitudes which involve at least one virtual gluon can be deduced fairly easily, as mentioned in the last subsection \ref{subsec:VirtualAmps}. This is  possible if the squared amplitude has a compact analytical form analogous to that at previous orders, which it partially does.

The all-gluons-real five-loops amplitude squared may be cast in the usual compact form:
\begin{align}
\cW^{\R\R\R\R\R}_{12345} = \cWR^{\R\R\R\R\R}_{12345} + \cWI^{\R\R\R\R\R}_{12345} \,, \label{eq:EikAmp_5Loop}
\end{align}
where the reducible contribution is given by:
\begin{align}
&\cWR^{\R\R\R\R\R}_{12345} = \frac{1}{s_5} \sum_{ijk\ell m=1}^5 \left[\cW^\R_i \cW^{\R\R\R\R}_{jk\ell m} + \cW^{\R\R}_{ij} \cW^{\R\R\R}_{k\ell m}  \right]\notag\\
=& \prod_{i=1}^5 \cW^\R_i + \sum_{ijk\ell m=1}^5 \Big[ \cW^\R_i \,\cW^\R_j \,\cW^\R_k\, \cWI^{\R\R}_{\ell m}\, \Theta^{i<j<k} +\notag\\
&+ \cW^\R_i\, \cW^\R_j \,\cWI^{\R\R\R}_{k\ell m} \,\Theta^{i<j} + \cW^\R_i\, \cWI^{\R\R\R\R}_{jk\ell m}+\notag \\
&+ \cW^\R_i\, \cWI^{\R\R}_{jk}\, \cWI^{\R\R}_{\ell m}\, \Theta^{j<\ell} + \cWI^{\R\R}_{ij}\, \cWI^{\R\R\R}_{k\ell m}\Big]\,.
\label{eq:EikAmp_5Loop_Red}
\end{align}
Note that this reducible contribution contains finite-$N_c$ terms coming from the four-loops irreducible part $\cWI^{\R\R\R\R}_{jk\ell m}$. The irreducible contribution $\cWI^{\R\R\R\R\R}_{12345}$ reads:
\begin{align}
\cWI^{\R\R\R\R\R}_{12345} = \cWIR^{\R\R\R\R\R}_{12345} + \cWII^{\R\R\R\R\R}_{12345} \,, \label{eq:EikAmp_5Loop_Irred}
\end{align}
where the term $\cWIR^{\R\R\R\R\R}_{12345}$, which resembles previous orders, is given by:
\begin{align}
&\cWIR^{\R\R\R\R\R}_{12345}=\,2\,\CF\,\CAfour\left[ \cD^{12345}_{ab}+\cA^{12}_{ab} \prod_{i=3}^5 \cAb^{1i}_{ab} +\right.\notag \\
&+ \sum_{jk\ell m=2}^5 \left(\cA^{1j}_{ab}\, \cAb^{1k}_{ab}\, \cBb^{1\ell m}_{ab} \,\Theta^{j<k} + \cA^{1j}_{ab}\, \cCb^{1k\ell m}_{ab} + \right.  \notag \\
&\left. + \cB^{1jk}_{ab} \, \cBb^{1\ell m}_{ab}\, \Theta^{j<\ell} + \cA_{ab}^{1j}\, \fAb_{ab}^{1k\ell m}\, \Theta^{k<\ell < m} \right)+  \notag \\
&\left.+\sum_{k\ell m=3}^5 \widetilde{\fA}_{ab}^{12k\ell m}\,\Theta^{\ell<m}+ \mathbb{A}_{ab}^{12345} +  \fB^{12345}_{ab} \right]+ \notag\\
&+4\,\CF \,\CAcub\left(\CF-\frac{\CA}{2}\right)\sum_{jk\ell m=2}^5\cAb_{ab}^{1j}\,\bA_{ab}^{1k\ell m}\,\Theta^{k<\ell < m} \,,  \label{eq:EikAmp_5Loop_Irred_Red}
\end{align}
with $\fAb_{ab}^{ik\ell m} = \fA_{ab}^{ik\ell m}/ w_{ab}^i$, and the pseudo-antennae $\widetilde{\fA}$, $\bbA$ and $\fB$ are generalizations of the previous-order pseudo-antenna \eqref{eq:Pseudo-Antennas_Defs0} (just as $\cB$ is a generalization of the two-loops antenna $\cA$). They read:
\begin{subequations}
\begin{align}
\widetilde{\fA}_{ab}^{ijk\ell m} &= w_{ab}^i \Big(\cA_{ai}^{jk} \cBb_{ai}^{j\ell m} + \cA_{ib}^{jk} \cBb_{ib}^{j\ell m } - \cA_{ab}^{jk} \cBb_{ab}^{j\ell m} \Big),\\
\bbA_{ab}^{ijk\ell m} &= w_{ab}^i \left(\cA_{ai}^{jk} \cAb_{ai}^{j\ell} \cAb_{ai}^{jm} + \cA_{ib}^{jk} \cAb_{ib}^{j\ell} \cAb_{ib}^{jm} - \right. \notag \\
 & \hspace{108pt}  \left. - \cA_{ab}^{jk} \cAb_{ab}^{j\ell} \cAb_{ab}^{jm} \right),   \\
 \fB^{ijk\ell m}_{ab} &= w_{ab}^i \left(\fA_{ai}^{jk\ell m} + \fA_{ib}^{jk\ell m} - \fA_{ab}^{jk\ell m} \right).
\end{align}
\end{subequations}
The totally-irreducible term which represents the new genuine finite-$N_c$ contribution at five loops reads:
\begin{align}
\cWII^{\R\R\R\R\R}_{12345} = 4\, \CF \, \CAcub \left(\CF-\frac{\CA}{2} \right) \cN^{\R\R\R\R\R}_{12345}\,,
\label{eq:EikAmp_5Loop_Irred_Irred}
\end{align}
where $\cN^{\X}_{12345}$ is a function containing the angular part of the finite-$N_c$ contribution for configuration $X$.

There are $8$ out of the $16$ squared amplitudes mentioned above that do actually contain mirror/Bose-symmetry breaking terms. These correspond to configurations in which both gluons $k_1$ and $k_2$ are real $\cW^{\R\R \mr{x}_3 \mr{x}_4 \R}_{12345}$, as well as $\cW^{\R\V\R\R\R}_{12345}$, $\cW^{\R\V\R\V\R}_{12345}$, $\cW^{\V\R\R\R\R}_{12345}$ and $\cW^{\V\R\R\V\R}_{12345}$. While the four squared amplitudes $\cW^{\R\R \mr{x}_3 \mr{x}_4 \R}_{12345}$ contain ``new'' totally-irreducible finite-$N_c$ contributions $\cN_{12345}^{\R\R \mr{x}_3\mr{x}_4\R}$, the other four amplitudes squared contain finite-$N_c$ terms related to the four-loops function $\cN_{jk\ell m}^{\R\R\mr{x}_\ell\R}$ in their reducible and/or irreducible parts (see Eqs. \eqref{eq:EikAmp_5Loop_Red} and \eqref{eq:EikAmp_5Loop_Irred_Red}), which itself breaks mirror and Bose symmetry.  The remaining $8$ squared amplitudes not mentioned above are perfectly symmetric under the interchange of any pair of emitters and are Bose symmetric.

Due to the mirror-symmetry breaking shared by the squared amplitudes $\cW^{\R\R \mr{x}_3 \mr{x}_4 \R}_{12345}$ it has not been possible to write their totally-irreducible parts $\cN^{\R\R \mr{x}_3 \mr{x}_4 \R}_{12345}$ in a compact analytical form as was the case at four loops. We provide the full expressions of these functions in the accompanying {\tt Mathematica} notebook ``N.nb''. All other squared amplitudes have a vanishing totally-irreducible contribution $\cN^{\X}_{12345}$.

In fact for all configurations in which either gluon $k_1$ or $k_2$ is virtual we have the following compact form of the squared amplitudes:
\begin{subequations}
\begin{align}
\cW^{\R\V \mr{x}_3\mr{x}_4 \R}_{12345} &= \frac{\cW_{12}^{\R\V}}{\cW_1^\R}\,\cW^{\R\mr{x}_3\mr{x}_4 \R}_{1345}   \,, \\
\cW^{\V\mr{x}_2\mr{x}_3\mr{x}_4 \R}_{12345}  & = \cW^\V_{1} \,\cW^{\mr{x}_2\mr{x}_3\mr{x}_4 \R}_{2345}\,.
\end{align}
\end{subequations}

We note that, as stated before, the calculation of the virtual-emission squared amplitudes involves attaching each virtual gluon one time in the bra-side of the squared amplitude and another in the ket-side (see Eqs. \eqref{eq:EikAmp_VirTreat1} and \eqref{eq:EikAmp_VirTreat2}). Due to symmetry reasons, and for the case of a single virtual gluon such as $\cW^{\R\R\R\V\R}_{12345}$, it is sufficient to only consider the attachment of the virtual gluon in one side and multiply by a factor of 2. Furthermore, and for the emission of two or more virtual gluons, it turns out that for squared amplitudes \emph{not} involving totally-irreducible finite-$N_c$ contributions, it is also sufficient to consider the attachment of the virtual gluons just in one side and multiply by a factor of 2 for each virtual gluon. This is true for configurations like $\cW^{\R\V\V\R}_{1234}$ at four loops and $\cW^{\R\V\R\V\R}_{12345}$ at five loops since they are free of (new) finite-$N_c$ contributions, i.e., free of the terms $\cN_{1234}^\X$ at four loops and $\cN_{12345}^\X$ at five loops.

However, for the squared amplitude $\cW^{\R\R\V\V\R}_{12345}$ one must explicitly include the attachment of virtual gluons in each side. By symmetry it is still sufficient to consider the attachment of one of the virtual gluons in one side but the other virtual gluon must be attached in both sides. The difference between attaching the second virtual gluon in one side and another results in a finite-$N_c$ contribution which affects the function $\cN_{12345}^{\R\R\V\V\R}$. This is ultimately related to the breaking of the mirror symmetry which occurs for diagrams with gluons $k_1$ and $k_2$ real.

We discuss the Bose and mirror symmetries of the five-loops amplitude squared and the source of their breakings in the coming two subsections.

\subsubsection{Bose symmetry}
\label{subsec:BoseSymmetry}

Analogous to the previous order, Eq. \eqref{eq:EikAmp_4loop_Lnc-Fnc_Facorised}, the five-loops amplitude squared \eqref{eq:EikAmp_5Loop} can be split into large- and finite-$N_c$ contributions:
\begin{align}
\cW^{\R\R\R\R\R}_{12345} = \cW^{\R\R\R\R\R,\,\mr{lN_c}}_{12345} + \cW^{\R\R\R\R\R, \,\mr{fN_c}}_{12345}\,, \label{eq:EikAmp_5loop_Lnc-Fnc_Facorised}
\end{align}
where:
\begin{multline}
\cW^{\R\R\R\R\R, \,\mr{fN_c}}_{12345} = \cWII^{\R\R\R\R\R}_{12345} + \sum_{ijk\ell m=1}^5 \cW^\R_i \, \cWII^{\R\R\R\R}_{jk\ell m} +  \\
 +4\,\CF \,\CAcub\left(\CF-\frac{\CA}{2}\right)\sum_{jk\ell m=2}^5\cAb_{ab}^{1j} \, \bA_{ab}^{1k\ell m}\,\Theta^{k<\ell < m}\,.
\label{eq:EikAmp_5loop_Fnc}
\end{multline}
Unlike the finite-$N_c$ contribution, we have the full compact analytical form of the large-$N_c$ contribution as confirmed via comparison with the output of the {\tt EikAmp} program. The latter contribution is totally Bose symmetric under permutations of the gluons. The reducible term \eqref{eq:EikAmp_5Loop_Red} is totally Bose symmetric after subtracting off the four-loops finite-$N_c$ contribution, i.e., after the replacement $\cWI^{\R\R\R\R}_{jk\ell m} \to \cWIR^{\R\R\R\R}_{jk\ell m}$ in Eq. \eqref{eq:EikAmp_5Loop_Red}.

\subsubsection{Mirror-symmetry breaking}
\label{subsec:MirrorSymmetry}

As previously stated the mirror-symmetry-breaking contributions only exist for configurations in which at least two out of the gluons $\{k_1,k_2,k_3\}$ are real regardless of the rest of the gluons. All other squared amplitudes are free of such terms. Moreover, the squared amplitudes for all configurations are symmetric under the permutation of the legs of the dipoles emitting the hardest and softest gluons. That is, if the amplitude squared for a given configuration $X$ is given by:
\begin{align}
\cW^{\X}_{12345}  \propto \mathcal{C}^{ai_2 i_3 i_4 i_5}_{b j_2 j_3 j_4 j_5} \, w_{ab}^1 w_{i_2 j_2}^2 w_{i_3 j_3}^3 w_{i_4 j_4}^4 w_{i_5 j_5}^5\,,
\label{eq:Mirror_Sym_A}
\end{align}
where $\mathcal{C}^{a \hdots i_5}_{b\hdots j_5}$ is the corresponding color factor for the specific Feynman diagram considered, then permuting the legs $(a\leftrightarrow b)$ and/or $(i_5 \leftrightarrow j_5)$ would result in an identical expression, or more specifically an identical color factor, to that in \eqref{eq:Mirror_Sym_A}, i.e.:
\begin{align}
\text{Eq. \eqref{eq:Mirror_Sym_A}} &=  \mathcal{C}^{\tcr{b} i_2 i_3 i_4 i_5}_{\tcr{a} j_2 j_3 j_4 j_5} \, w_{\tcr{ba}}^1 w_{i_2 j_2}^2 w_{i_3 j_3}^3 w_{i_4 j_4}^4 w_{i_5 j_5}^5  \notag \\
 &= \mathcal{C}^{a i_2 i_3 i_4 \tcr{j_5}}_{b j_2 j_3 j_4 \tcr{i_5}}  \, w_{ab}^1 w_{i_2 j_2}^2 w_{i_3 j_3}^3 w_{i_4 j_4}^4 w_{\tcr{j_5 i_5}}^5 \notag \\
  &= \mathcal{C}^{\tcr{b} i_2 i_3 i_4 \tcr{j_5}}_{\tcr{a} j_2 j_3 j_4 \tcr{i_5}}  \, w_{\tcr{ba}}^1 w_{i_2 j_2}^2 w_{i_3 j_3}^3 w_{i_4 j_4}^4 w_{\tcr{j_5 i_5}}^5\,.
 \label{eq:Mirror_Sym_B}
\end{align}
The difference between, say, the antennae $w^1_{ab}$ and $w^1_{ba}$ is that for $w^1_{ab}$ gluon $k_1$ is  attached to leg $a$ in the ket amplitude and to leg $b$ in the bra amplitude, whereas for $w^1_{ba}$ the attachment is reversed. The permutation of the legs of dipoles emitting gluons other than the softest and hardest does not however result in an identical color factor, and this is what we refer to as ``mirror-symmetry breaking''. Up to three loops these permutations of the emitting legs had no effect on the form of the eikonal amplitudes squared. This is, however, no longer the case at four and five loops (and perhaps at higher loops too).

\begin{figure}[ht]
\centering
\includegraphics[width=.48\textwidth]{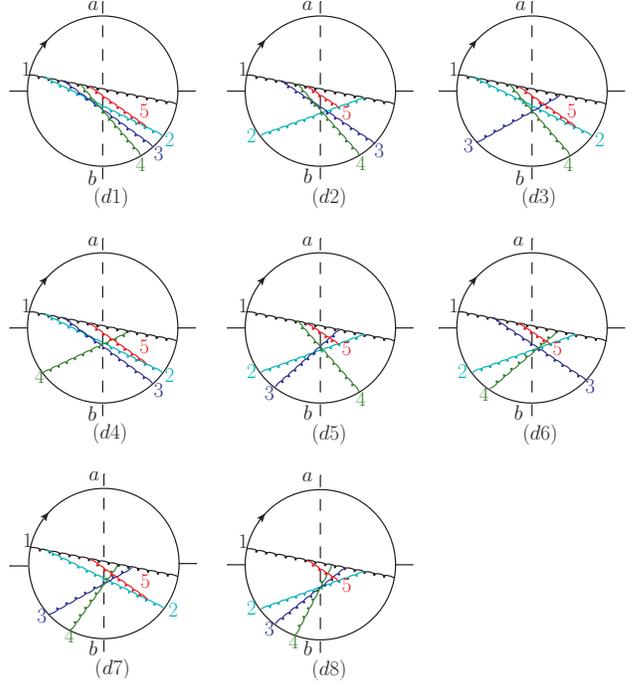}
\caption{An example of Feynman diagrams, for the amplitude squared, which are not symmetric under the interchange of one or more pairs of emitters of gluons $k_2$, $k_3$ and $k_4$.\label{fig:EikAmp_5loop_SymBreak}}
\end{figure}

In order to illustrate the origin of the said non-mirror-symmetric contributions diagrammatically let us consider, for instance, the  Feynman diagrams depicted in Fig. \ref{fig:EikAmp_5loop_SymBreak},
contributing to the all-gluons-real amplitude squared $\cW^{\R\R\R\R\R}_{12345}$. Excluding permutations of the emitters of $k_1$ and $k_5$, under which the color factor is symmetric as stated above, there are $8$ possible permutations of the emitters of the remaining gluons $k_2$, $k_3$ and $k_4$. These are represented in diagrams $(d1)$ -- $(d8)$ of Fig. \ref{fig:EikAmp_5loop_SymBreak}. Their corresponding contributions to the amplitude squared are given by:
\begin{subequations}
\begin{alignat}{2}
(d1):\,&& -\frac{3\,\CF\,\CAcub}{8} &\left(\CF-\frac{\CA}{2}\right) w^1_{ab} w^2_{1b} w^3_{1b} w^4_{1b} w^5_{12}\,,\\
(d2):\,&&    -\frac{\CF\,\CAcub}{4} &\left(\CF-\frac{\CA}{2}\right) w^1_{ab} w^2_{\tcr{b1}} w^3_{1b} w^4_{1b} w^5_{12}\,,\\
(d3):\,&&    +\frac{\CF\,\CAcub}{4} &\left(\CF-\frac{\CA}{2}\right) w^1_{ab} w^2_{1b} w^3_{\tcr{b1}} w^4_{1b} w^5_{12}\,,\\
(d4):\,&&    +\frac{\CF\,\CAcub}{4} &\left(\CF-\frac{\CA}{2}\right) w^1_{ab} w^2_{1b} w^3_{1b} w^4_{\tcr{b1}} w^5_{12}\,.
\end{alignat}\label{eq:EikAmp_5Loop_RRRRR-SymBreak}
\end{subequations}
The other diagrams $(d5)$, $(d6)$, $(d7)$ and $(d8)$, correspond to color factors that are identical to those of $(d3)$, $(d4)$, $(d2)$ and $(d1)$, respectively. The lack of this (mirror) symmetry makes it quite difficult to find a compact analytical form for the totally-irreducible contributions $\cN_{12345}^{\X}$ to the five-loops amplitude squared.

We finally note that adding up the four functions $ \cN^{\X}_{12345}$ one finds:
\begin{align}
\cN^{\R\R\R\R\R}_{12345} + \cN^{\R\R\R\V\R}_{12345} + \cN^{\R\R\V\R\R}_{12345} + \cN^{\R\R\V\V\R}_{12345} = 0\,.
\end{align}
This equality has been deduced from the output of the {\tt EikAmp} program.

\subsubsection{Exponentiation}
\label{sec:Exponentiation}

First, it is intriguing to notice that the three-, four- and five-loops eikonal amplitudes squared, both reducible and irreducible contributions, are written in terms of the two-loops antenna $\cA_{ij}^{k\ell}$, including the pseudo-antenna function $\bA_{ab}^{ijk\ell}$ \eqref{eq:Pseudo-Antennas_Defs} as is clear from the relations between the $t/u$-channel functions and the two-loops antenna given in Eq. \eqref{eq:T-U_Channel_Funs}. Therefore the two-loops amplitude squared, and the antenna $\cA_{ij}^{k\ell}$ in particular, may be considered as a \emph{web} \cite{Laenen:2010uz} since it represents a two-eikonal  irreducible diagram.\footnote{The two eikonal lines are the quark and anti-quark lines.}

Secondly, Eqs. \eqref{eq:EikAmp_2Loop}, \eqref{eq:EikAmp_2Loop_Irred}, \eqref{eq:EikAmp_3Loop}, \eqref{eq:EikAmp_3Loop_Irred00}, \eqref{eq:EikAmp_4Loop_Red}, \eqref{eq:EikAmp_4Loop_Irred_Red}, \eqref{eq:EikAmp_5Loop_Red} and \eqref{eq:EikAmp_5Loop_Irred_Red} seem to suggest a pattern of exponentiation for the eikonal amplitude squared (in agreement with the findings of Refs. \cite{Sterman:1981jc, Gatheral:1983cz, Frenkel:1984pz, Catani:1984dp}).

\section{Conclusions}
\label{sec:Conclusion}

In the present paper we have considered the computation of QCD scattering amplitudes squared in the special eikonal approximation. In the said approximation, radiated gluons are assumed to be soft. If the gluons are additionally assumed to be ordered in energy then the calculations are significantly simplified and it has thus been possible to go beyond two loops at finite $N_c$ for the first time in literature, for the simple process $e^+e^- \to q \bar{q}$ accompanied with the emission of soft gluons.

First, we have shown how the eikonal approximation follows from standard QCD calculations based on Feynman rules when radiated gluons have energies that are much less than the hard scale of the process. The said Feynman rules are then replaced by the more simplified eikonal rules. The latter rules have made calculations at higher orders in perturbation theory very accessible.  Moreover, assuming on-mass-shell limit we have demonstrated through explicit calculations that virtual corrections in the eikonal limit reduce to simply be minus their corresponding real-emission contribution (for the softest gluon). This led to even more simplifications in the required calculations at a given order.

Secondly, starting with the amplitude for the emission of the $m^{\mr{th}}$ gluon by an ensemble of $(m+1)$ partons and iterating down to the Born level, we have derived a general analytical form of the amplitude squared for the emission of any number of soft, energy-ordered gluons in $e^+e^-$ annihilation. With the aid of the {\tt ColorMath} package, the latter analytical form has been implemented into a {\tt Mathematica} program that is capable of computing the eikonal amplitude squared for the full set of Feynman diagrams at any given order. Furthermore, we have illustrated how one may easily find the eikonal amplitude squared in the large-$N_c$ limit with no further work, simply by setting $\CF \to \CA/2$ in the program.

For practical purposes we have presented the full eikonal squared amplitudes for the emission of up to five gluons in the final state in the process $e^+e^- \to q \qbar$, at finite $N_c$. Whilst the three-loops squared amplitude displayed no major form-wise departure from the well-known two-loops squared amplitude, the four- and five-loops squared amplitudes did. New antenna functions and genuine finite-$N_c$ corrections first appear at four loops. Unlike previous orders, at the fourth and fifth order the amplitude squared is manifestly non-symmetric under Bose permutations (of the emitted gluons). It is additionally non-symmetric under mirror permutations (the interchange of one or more pairs of emitters of gluons) other than for the hardest and softest gluons. This symmetry breaking has tremendously complicated the task of writing the amplitude squared at five loops in a compact form, as has been performed at previous orders. We expect the level of complexity to rise up as one moves to higher orders.

Nonetheless three points are noteworthy regarding eikonal amplitudes squared:
\begin{itemize}
\item At a given loop order all one has to calculate is the all-gluons-real squared amplitude. All other squared amplitudes, for all possible virtual configurations of gluons, can straightforwardly be deduced from the latter.
\item Higher-order squared amplitudes can be shown to be built in terms of the two-loops antenna function. The latter may be considered as a web.
\item The trend of the amplitudes squared, at least up to five loops, seems to suggest a pattern of exponentiation.
\end{itemize}

Lastly, it is worthwhile to recall that the eikonal approximation only guarantees the resummation of up to single logarithms for a given observable. To go one more logarithm down it is sufficient to consider the next-to-eikonal approximation, a task which seems to naturally follow our present work and which we shall treat in our coming publications. 

\acknowledgments

This work is supported in part by the CNEPRU Research Project D01320130009.

\appendix

\section{Eikonal Feynman rules}
\label{sec:app:EffectiveFeynmanRules}

\begin{figure}[ht]
\includegraphics[scale=.64]{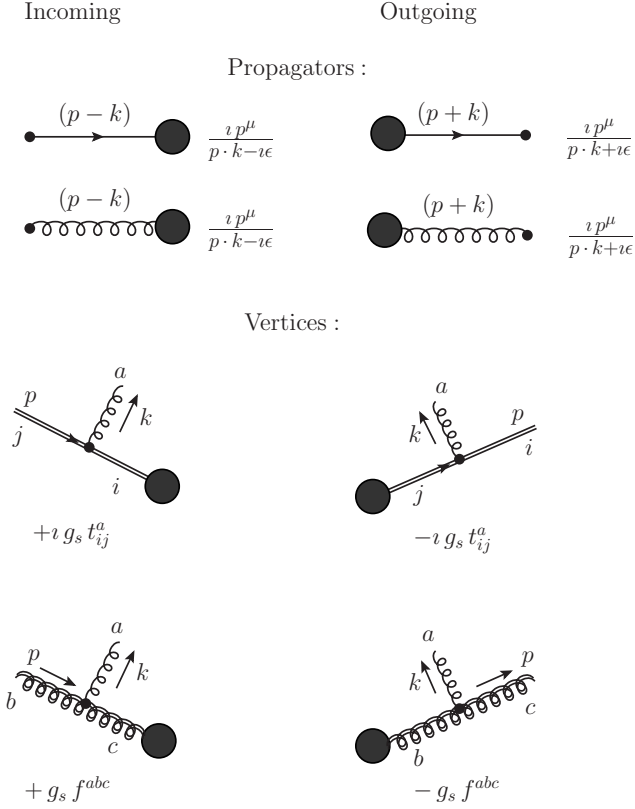}
\caption{Eikonal Feynman rules for QCD. The ``double'' lines represent the radiating quark and gluon eikonal lines.\label{fig:EikFeynRules}}
\end{figure}

The effective eikonal Feynman rules (in the Feynman gauge) are depicted in Fig. \ref{fig:EikFeynRules}.
In the latter figure we only display the quark and gluon eikonal lines. The corresponding rules for the anti-quark eikonal line are identical to those for the quark line except that one multiplies the vertices by $(-1)$.

The sign convention adopted herein for the eikonal vertex is as follows \cite{Catani:1996jh, Catani:1996vz}: in Eq. \eqref{eq:EikAmp_general} the color operator $\bT^a_i$ for the emission of gluon $a$ off a harder parton $i$ is replaced by:
\begin{itemize}
\item $+ \bt^a = + t^a_{ij}$, the generator in the fundamental representation, if $i$ is an outgoing (incoming) quark (anti-quark).
\item $-\bt^a = - t^a_{ij}$ if $i$ is an incoming (outgoing) quark (anti-quark).
\item $+(-)\,\imath \bbf^a = +(-) \imath f^a_{bc}$, the generator in the adjoint representation, if $i$ is an outgoing (incoming) gluon.
\end{itemize}

Note that in Fig. \ref{fig:EikFeynRules} the minus sign in the denominator of the eikonal propagator (for incoming partons) has been absorbed into the corresponding vertex.

Moreover, it is worth noting that Fig. \ref{fig:EikFeynRules} includes only diagrams for the \emph{emission} of a soft gluon, i.e., the direction of the gluon's four-momentum is away from the vertex. For the \emph{absorption} of a soft gluon, i.e., the direction of the gluon's four-momentum is towards the vertex, one multiplies the corresponding vertices by $(-1)$. This minus sign then cancels against the minus sign from the propagator of the parton absorbing the gluon, since the latter's momentum is $(p-k)$ (see Fig. \ref{fig:EikFeynRules}).

\bibliography{EikAmp_refs}

\end{document}